\documentclass[pra,showpacs,floatfix,nofootinbib]{revtex4}
\usepackage{amsmath}
\usepackage{amssymb}
\usepackage{amsthm}
\usepackage{amscd}
\usepackage{amsfonts}
\usepackage[dvips]{graphicx}
\usepackage{latexsym}
\usepackage{mathrsfs}
\usepackage{bm}
\usepackage{subfigure}
\usepackage{color}

\catcode`\@=11
\renewcommand\footnoterule{  \kern-3\p@
  \hrule\@width.4\columnwidth
  \kern2.6\p@}
\renewcommand\@makefntext[1]{    \parindent 1em\noindent
    \hb@xt@1.8em{\hss$^{\@thefnmark}$)}\hspace{2pt}    \footnotesize\rmfamily#1}  \def\@makefnmark{\hspace{.5pt}\hbox{$^{\@thefnmark}$\hspace{-1pt}}}
\renewcommand\footnoterule{  \kern-3\p@
  \hrule\@width.4\columnwidth
  \kern2.6\p@}
\renewcommand\@makefntext[1]{    \parindent 1em\noindent
    \hb@xt@1.8em{\hss$^{\@thefnmark}$)}\hspace{2pt}    \footnotesize\rmfamily#1}  \def\@makefnmark{\hspace{.5pt}\hbox{$^{\@thefnmark}$\hspace{-1pt}}}
\renewcommand\footnoterule{  \kern-3\p@
  \hrule\@width.4\columnwidth
  \kern2.6\p@}
\renewcommand\@makefntext[1]{    \parindent 1em\noindent
    \hb@xt@1.8em{\hss$^{\@thefnmark}$)}\hspace{2pt}    \footnotesize\rmfamily#1}  \def\@makefnmark{\hspace{.5pt}\hbox{$^{\@thefnmark}$\hspace{-1pt}}}
\renewcommand\footnoterule{  \kern-3\p@
  \hrule\@width.4\columnwidth
  \kern2.6\p@}
\renewcommand\@makefntext[1]{    \parindent 1em\noindent
    \hb@xt@1.8em{\hss$^{\@thefnmark}$)}\hspace{2pt}    \footnotesize\rmfamily#1}  \def\@makefnmark{\hspace{.5pt}\hbox{$^{\@thefnmark}$\hspace{-1pt}}}
\renewcommand\footnoterule{  \kern-3\p@
  \hrule\@width.4\columnwidth
  \kern2.6\p@}
\renewcommand\@makefntext[1]{    \parindent 1em\noindent
    \hb@xt@1.8em{\hss$^{\@thefnmark}$)}\hspace{2pt}    \footnotesize\rmfamily#1}  \def\@makefnmark{\hspace{.5pt}\hbox{$^{\@thefnmark}$\hspace{-1pt}}} 
\def\RR{\mathbb{R}}

\newcommand{\cC}{\mathcal{C}}

\newcommand{\cP}{\mathcal{P}}
\newcommand{\cT}{\mathcal{T}}

\newcommand{\bu}{\mathbf{u}}

\newcommand{\bT}{\mathbf{T}}

\newcommand{\be}[1]{\begin{equation}\label{#1}}
\newcommand{\ee}{\end{equation}}
\newcommand{\ba}[1]{\begin{eqnarray}\label{#1}}
\newcommand{\ea}{\end{eqnarray}}
\newcommand{\rf}[1]{(\ref{#1})}

\begin{document}

\title{Nonlinear $\mathcal{P}\mathcal{T}-$symmetric plaquettes}
\author{Kai Li}
\affiliation{Department of of Mathematics and Statistics, University of
Massachusetts,\\
Amherst, MA 01003-9305, USA}
\author{P. G. Kevrekidis}
\affiliation{Department of of Mathematics and Statistics, University of
Massachusetts,\\
Amherst, MA 01003-9305, USA}
\author{Boris A. Malomed}
\affiliation{
Department of Physical Electronics, School of Electrical Engineering,
Faculty of Engineering, \\
Tel Aviv University, Tel Aviv 69978, Israel}
\author{Uwe G{\"u}nther}
\affiliation{Helmholtz Center Dresden-Rossendorf,  POB 510119,\\ D-01314
Dresden, Germany}

\begin{abstract}
We introduce four basic two-dimensional (2D) plaquette configurations with
onsite cubic nonlinearities, which may be used as building blocks for 2D $%
\mathcal{PT}$-symmetric lattices. For each configuration, we develop a
dynamical model and examine its $\mathcal{P}\mathcal{T}\ $symmetry. The
corresponding nonlinear modes are analyzed starting from the Hamiltonian
limit, with zero value of the gain-loss coefficient, $\gamma $. Once the
relevant waveforms have been identified (chiefly, in an analytical form),
their stability is examined by means of linearization in the vicinity
of stationary points. This reveals diverse and, occasionally, fairly complex
bifurcations. The evolution of unstable modes is explored by means of direct
simulations. In particular, stable localized modes are found in these
systems, although the majority of identified solutions is unstable.
\end{abstract}


\maketitle

\everymath{\displaystyle}

\section{Introduction}

The theme of $\mathcal{PT}$ (parity--time) symmetric systems was initiated
in the works of Bender and collaborators~\cite{bend} as an alternative to
the standard quantum theory, where the Hamiltonian is postulated to be
Hermitian. The principal conclusion of these works was that $\mathcal{PT}$%
-invariant Hamiltonians, which are not necessarily Hermitian, may still give
rise to completely real spectra, thus being appropriate for the description
of physical settings. In terms of the Schr{\"{o}}dinger-type Hamiltonians,
which include the usual kinetic-energy operator and the potential term, $%
V(x) $, the $\mathcal{PT}$-invariance admits complex potentials, subject to
constraint that $V^{\ast }(x)=V(-x)$.

Recent developments in optics have resulted in an experimental realization
of the originally theoretical concept of the $\mathcal{PT}$-symmetric
Hamiltonians, chiefly due to the work by Christodoulides and co-workers~\cite%
{christo1} (see also \cite{haifa-prl-2008}). It has been demonstrated that
the controllable imposition of symmetrically set and globally balanced gain
and loss may render optical waveguiding arrays a fertile territory for the
construction of $\mathcal{PT}$-symmetric complex potentials. The first two
such realizations made use of couplers composed of two waveguides with and
without loss~\cite{salamo} (so-called passive $\mathcal{PT}-$couplers), or,
in more ``standard" form, a pair of coupled waveguides, one carrying gain
and the other one loss~\cite{kip}. In fact, more general models of linearly
coupled active (gain-carrying) and passive (lossy) intrinsically nonlinear
waveguides, without imposing the condition of the gain-loss balance, were
considered earlier, and stable solitons were found in them \cite{Winful},
including exact solutions \cite{Atai} (see also a brief review in Ref. \cite%
{Chaos}). Recently, an electronic analog of such settings has also been
implemented~\cite{tsampikos_recent,tsampikos_recent2}. Configurations with a
hidden $\mathcal{P}\mathcal{T}$ symmetry have been identified also in
fine-tuned parameter regions of microwave billiards \cite{darmstadt-pt}.
Effects of the nonlinearity in a Gross-Pitaevski equation on the $\cP\cT$ properties of a Bose-Einstein condensate have been analyzed in \cite{cart-wun2012}. The
possibility to engineer $\mathcal{PT}$-symmetric \textit{oligomers}~(coupled
complexes of a few loss-and gain-carrying elements) \cite{pgk}, which may
include nonlinearity, was an incentive to a broad array of additional
studies on both the few-site systems and entire $\mathcal{PT}$-symmetric
lattices~\cite{kot1,sukh1,kot2,grae1,grae2,kot3,dmitriev1,dmitriev2}. More
recently, nonlinear $\mathcal{PT}$-symmetric systems, incorporating $%
\mathcal{PT}$-balanced nonlinear terms, have drawn considerable interest~too
\cite{miron}-\cite{YJH}.

\begin{figure}[tph]
\subfigure[\ mode 0+0-]{\scalebox{0.21}{\includegraphics{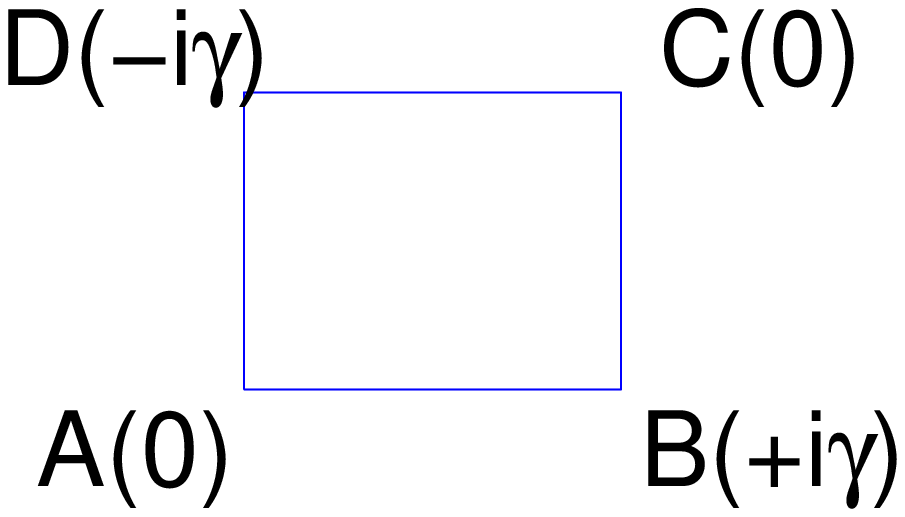}}} %
\subfigure[\ mode +-+-]{\scalebox{0.21}{\includegraphics{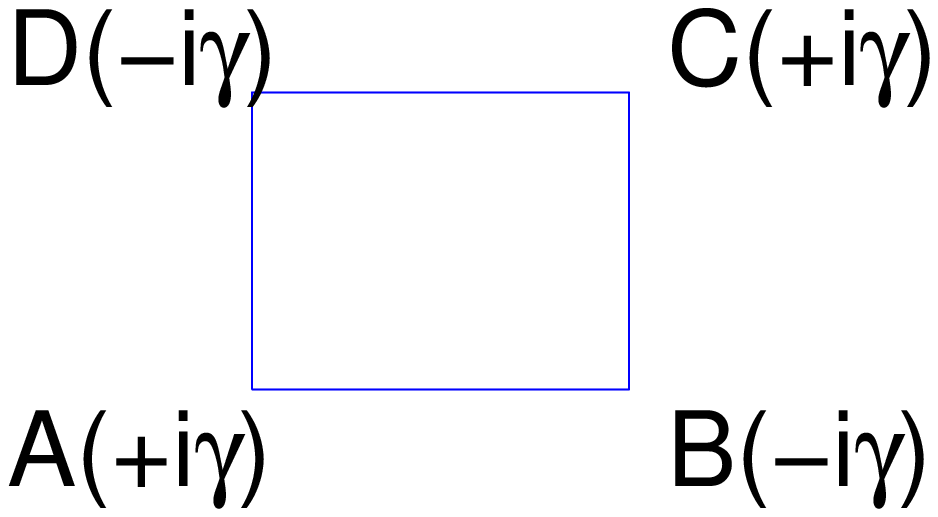}}} %
\subfigure[\ mode ++- -]{\scalebox{0.21}{\includegraphics{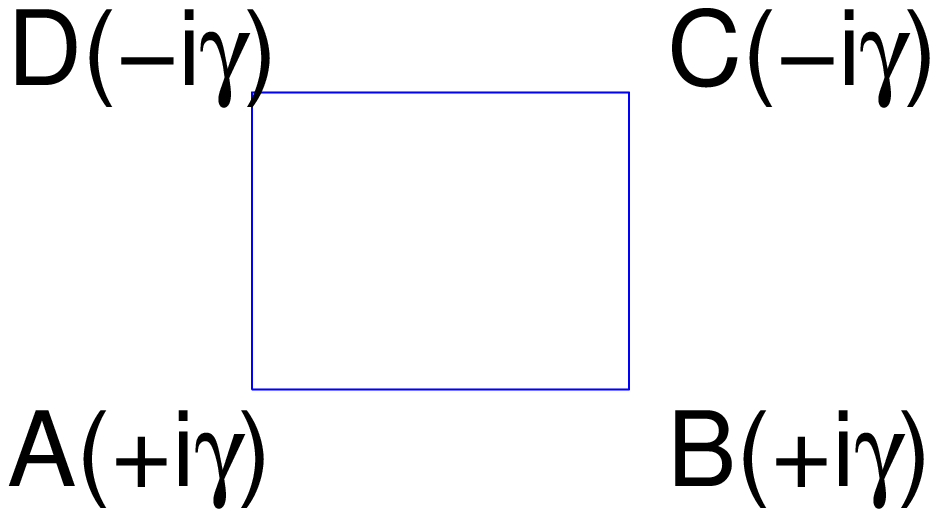}}} %
\subfigure[\ mode +-0+-]{\scalebox{0.21}{\includegraphics{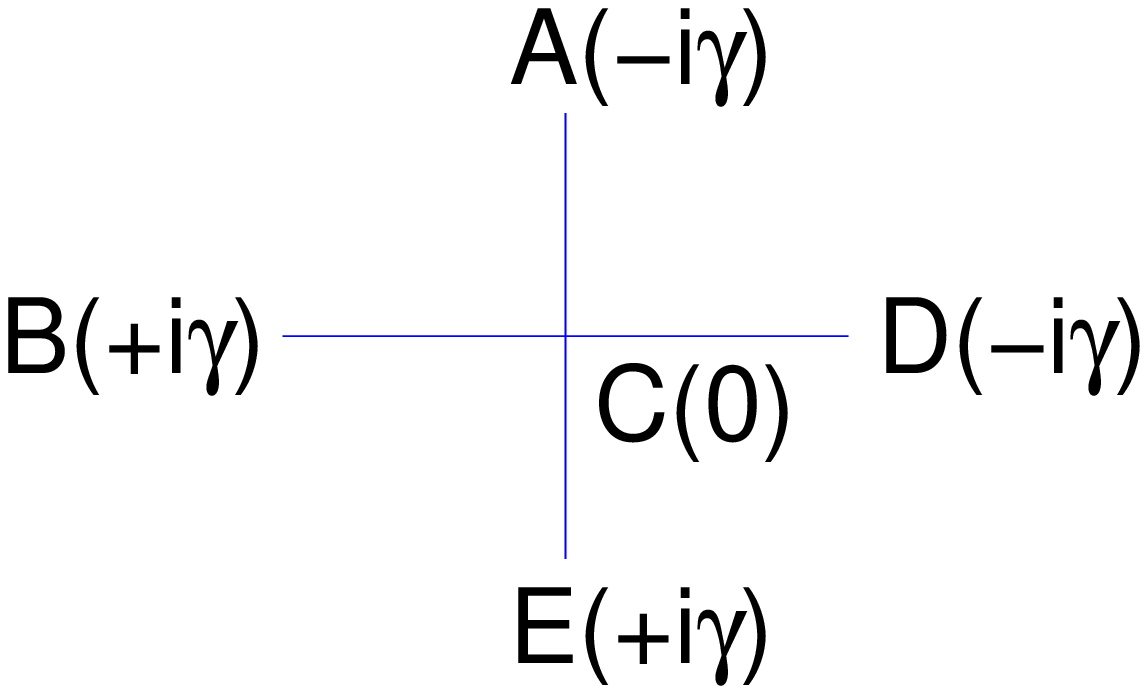}}}
\caption{(Color online) The different fundamental plaquette configurations
(i.e., two-dimensional \textit{oligomers}) including the linear balanced
gain and loss. Among these, (a), (c) and (d) are $\mathcal{PT}$-symmetric,
while (b) is not in the strict sense, but it is interesting too, as an
implementation of alternating gain and loss nodes in the plaquette pattern.
The nodes are labeled so as to connect the gain-loss profiles to the
evolution of individual nodes in dynamical simulations. The sets are coded
by chains of symbols, with $+,-$ and $0$ corresponding, respectively, to the
linear gain, loss, or absence of either effect at particular sites.}
\label{modes}
\end{figure}

Most of the $\mathcal{PT}$-invariant systems considered thus far have been
one-dimensional (1D) in their nature, although the stability of solitons in
2D periodic $\mathcal{PT}$-symmetric potentials has also been recently
investigated \cite{Yang}. Actually, 2D arrays of optical waveguides~can be
readily built \cite{Review} (the same is true about other quasi-discrete
systems, including electrical ones), hence, a natural
question is whether $\mathcal{PT}$-symmetric oligomers (and ultimately
lattices built of such building blocks) can be created in a 2D form. This
work aims to make a basic step in this direction, by introducing fundamental
2D plaquettes consisting, typically, of four sites (in one case, it will be
a five-site cross). These configurations, illustrated by Fig.~\ref{modes},
are inspired by earlier works on 2D Hamiltonian lattices described by
discrete nonlinear Schr{\"{o}}dinger equations~\cite{pgk_book}, where
diverse classes of modes, including discrete solitary vortices~\cite%
{malomed,pelin}, have been predicted and experimentally observed~\cite%
{moti,yuri}. The plaquettes proposed herein should be straightforwardly
accessible with current experimental techniques in nonlinear optics, as a
straightforward generalization of the coupler-based setting reported in Ref.~%
\cite{kip}. We start from the well-established Hamiltonian form of such
plaquettes in the conservative form, gradually turning on the gain-loss
parameter ($\gamma $), as the strength of the $\mathcal{PT}$-invariant
terms, to examine stationary states supported by the plaquettes, studying
their stability against small perturbations and verifying the results
through direct simulations. Actually, in this work we focus on those (quite
diverse, although, obviously, not most generic) modes that can be found in
an analytical form, while their stability is studied by means of numerical
methods. The analytical calculations and the manifestations of interesting
features, such as a potential persistence past the critical point of the
linear $\mathcal{PT}$ symmetry, are enabled by the enhanced symmetry of the
modes that we consider below. It is conceivable that additional asymmetric
modes may exist too within these 2D configurations.

Our principal motivation for studying the above systems stems from the
fact that realizations of $\mathcal{PT}$-symmetry e.g. within the
realm of nonlinear optics will be inherently endowed with nonlinearity.
Hence, it is only natural to inquire about the interplay of the above
type of linear systems with the presence of nonlinear effects.
In addition to this physical argument, there exists an intriguing
mathematical one which concerns the existence, stability and dynamical
fate of the nonlinear states in the presence of  $\mathcal{PT}$-symmetric
perturbations. In particular, previous works~\cite{pgk,grae2,konorecent4,cart-wun2012,grae3}
point to the direction
that neither the existence, nor the stability of $\mathcal{PT}$-symmetric
nonlinear states mirrors that of their linear counterparts (or respects
the phase transition of the latter generically).
The presentation of our results
is structured as follows. Section \ref{symmetry} contains a
part of the analytical results, including a detailed analysis of the $%
\mathcal{P}\mathcal{T}-$symmetry properties of the nonlinear Schr\"{o}dinger
type model, as well as the spectral properties of the linear Hamiltonian
subsystems. Section \ref{numerics} is devoted to the existence, stability
and dynamics of stationary modes in the nonlinear systems. Beside analytical
results, it contains a detailed presentation of the numerical findings. In
section \ref{conclu} we summarize conclusions and discuss directions for
future studies.

\section{The setup and symmetry properties\label{symmetry}}

\subsection{General techniques}

The dynamics of the 2D plaquettes that we are going to consider is described
by a multicomponent nonlinear Schr\"{o}dinger equation (NLSE)
\begin{equation}
i\dot{\mathbf{u}}=H_{L}\mathbf{u}+H_{NL}(\mathbf{u})\mathbf{u}  \label{a1}
\end{equation}%
built over a transposition-symmetric linear $N\times N$ Hermitian matrix
Hamiltonian $H_{L}=H_{L}^{T}$ and an additional nonlinear $N\times N$ matrix
operator, $H_{NL}(\mathbf{u})=H_{NL}^{T}(\mathbf{u})$. To understand the
symmetry properties of this NLSE, we first analyze the associated linear
problem
\begin{equation}
i\dot{\mathbf{u}}=H_{L}\mathbf{u,}  \label{a2}
\end{equation}%
and check then whether the symmetry is preserved by the nonlinear term, $%
H_{NL}(\mathbf{u})\mathbf{u}$. The analysis can be built, in a part, on
techniques developed for other nonlinear dynamical systems with symmetry
preservation \cite%
{nonlin-1,nonlin-2,nonlin-3,nonlin-4,nonlin-5,nonlin-6,nonlin-7,nonlin-8}.

For the present setups, the time reversal operation $\mathbf{T}$ can be
defined as the combined action of a scalar-type complex conjugation $%
\mathcal{T}$, $\mathcal{T}^{2}=I$, and the sign change of time, $%
t\rightarrow -t$, in full accordance with Wigner's original work which
introduced these concepts \cite{wigner-book}. For the linear Schr\"{o}dinger
equation (\ref{a2}) and its solutions
\begin{eqnarray}
\mathbf{u}(t) &=&\sum_{n=1}^{N}e^{-iE_{n}t}\mathbf{u}_{n},  \notag
\label{a2-1} \\
H_{L}\mathbf{u}_{n} &=&E_{n}\mathbf{u}_{n},
\end{eqnarray}%
this implies
\begin{eqnarray}
\mathbf{T}(i\partial _{t}\mathbf{u}) &=&\mathbf{T}(H_{L}\mathbf{u}),  \notag
\label{a2-2} \\
i\partial _{t}\mathbf{T}(\mathbf{u}) &=&\bar{H}_{L}\mathbf{T}(\mathbf{u}),
\notag \\
\mathbf{T}\mathbf{u}(t) &=&\mathcal{T}\mathbf{u}(t)|_{t\rightarrow
-t}=\sum_{n=1}^{N}e^{-i\bar{E}_{n}t}\bar{\mathbf{u}}_{n}\,,
\end{eqnarray}%
where the overbar denotes complex conjugation. From the actual form of the
gain-loss arrangements in the 2D plaquettes we can conjecture the existence
of certain plaquette-dependent parity operators $\mathcal{P}$, with $%
\mathcal{P}^{2}=I$, which will render the Hamiltonians $\mathcal{P}\mathcal{T%
}-$symmetric, $[\mathcal{P}\mathcal{T},H_{L}]=0$. To find an explicit
representation of these parity operators $\mathcal{P}$, we use the
following ansatz,
\begin{equation}
\mathcal{P}\in \mathbb{R}^{N\times N},\qquad \lbrack \mathbf{T},\mathcal{P}%
]=0  \label{a2-3}
\end{equation}%
together with the pseudo-Hermiticity condition
\begin{equation}
H_{L}^{\dagger}=\mathcal{P}H_{L}\mathcal{P}.  \label{a2-4}
\end{equation}%
The latter follows trivially from Eq. (\ref{a2-3}), $[\mathcal{P}\mathcal{T}%
,H_{L}]=0$, and the transposition symmetry, $H_{L}=H_{L}^{T}$. These parity
operators $\mathcal{P}$ will be used to check whether the corresponding
nonlinear terms $H_{NL}(\mathbf{u})$ satisfy the same $\mathcal{P}\mathbf{T}%
- $symmetry. In contrast to linear setups with the $\mathcal{P}\mathbf{T}%
-$symmetry being either exact ($[\mathcal{P}\mathbf{T},H_{L}]=0$,\ $\mathcal{%
P}\mathbf{T}\mathbf{u}\propto \mathbf{u}$) or spontaneously broken ($[%
\mathcal{P}\mathbf{T},H_{L}]=0$,\ $\mathcal{P}\mathbf{T}\mathbf{u}%
\not\propto \mathbf{u}$), the nonlinear setups considered in the present
paper allow for sectors of exact $\mathcal{P}\mathbf{T}-$symmetry ($[%
\mathcal{P}\mathbf{T},H_{NL}(\mathbf{u})]=0$,\ $\mathcal{P}\mathbf{T}\mathbf{%
u}\propto \mathbf{u}$) and of broken $\mathcal{P}\mathbf{T}-$symmetry ($%
\mathcal{P}\mathbf{T}\mathbf{u}\not\propto \mathbf{u}$\quad $\Longrightarrow
$\quad $\lbrack \mathcal{P}\mathbf{T},H_{NL}(\mathbf{u})]\neq 0$), as it is
common for nonlinear $\mathcal{P}\mathcal{T}$-symmetric systems \cite{miron}-%
\cite{YJH}.

\subsection{$\mathcal{P}\mathbf{T}-$symmetry properties of 2D plaquettes}

We start from the 2D plaquette of 0+0- type depicted as configuration (a) in
Fig.~\ref{modes}. This plaquette has only two (diagonally opposite) nodes
carrying the gain and loss, while the other two nodes bear no such effects.
The corresponding dynamical equations for the amplitudes at the four sites
of this oligomer are
\begin{eqnarray}
i\dot{u}_{A} &=&-k(u_{B}+u_{D})-|u_{A}|^{2}u_{A},  \notag \\
i\dot{u}_{B} &=&-k(u_{A}+u_{C})-|u_{B}|^{2}u_{B}+i\gamma u_{B},  \notag \\
i\dot{u}_{C} &=&-k(u_{B}+u_{D})-|u_{C}|^{2}u_{C},  \notag \\
i\dot{u}_{D} &=&-k(u_{A}+u_{C})-|u_{D}|^{2}u_{D}-i\gamma u_{D,}
\label{zpzm1}
\end{eqnarray}%
where $\gamma \in \mathbb{R}$ is the above-mentioned gain-loss coefficient,
and $k\in \mathbb{R}$ is a real coupling constant. The nonlinearity
coefficients are scaled to be $1$ (we use time $t$ as the evolution
variable, although in the mathematically equivalent propagation equations
for optical waveguides $t$ has to be identified with the propagation
distance, $z$).

Denoting $\mathbf{u}:=(u_{A},u_{B},u_{C},u_{D})^{T}\in \mathbb{C}^{4}$, the
matrices $H_{L}$ and $H_{NL}(\mathbf{u})$ in (\ref{a1}) take the form of
\begin{eqnarray}
 H_{L} &=&\left(
\begin{array}{cccc}
0 & -k & 0 & -k \\
-k & i\gamma & -k & 0 \\
0 & -k & 0 & -k \\
-k & 0 & -k & -i\gamma
\end{array}%
\right) =H_{L}^{T},  \label{a3} \\
 &=&-k(I+\sigma _{x})\otimes \sigma _{x}+i\frac{\gamma }{2}\sigma
_{z}\otimes (I-\sigma _{z}),  \notag \\
 H_{NL}(\mathbf{u}) &=&-\left(
\begin{array}{cccc}
|u_{A}|^{2} & 0 & 0 & 0 \\
0 & |u_{B}|^{2} & 0 & 0 \\
0 & 0 & |u_{C}|^{2} & 0 \\
0 & 0 & 0 & |u_{D}|^{2}
\end{array}%
\right) =H_{NL}^{T}(\mathbf{u})=H_{NL}^{\dagger}(\mathbf{u}).\label{nl-1}
\end{eqnarray}%
To find the parity matrix $\mathcal{P}$ which renders the linear Hamiltonian
$\mathcal{P}\mathcal{T}-$symmetric, $[\mathcal{P}\mathcal{T},H_{L}]=0$, we
use the pseudo-Hermiticity condition (\ref{a2-4}) and notice that
\begin{eqnarray}
H_{L} &=&H_{L,0}+H_{L,1},  \label{a5} \\
H_{L,0} &=&-k(I+\sigma _{x})\otimes \sigma _{x}=H_{L,0}^{\dagger},  \notag \\
H_{L,1} &=&i\frac{\gamma }{2}\sigma _{z}\otimes (I-\sigma
_{z})=-H_{L,1}^{\dagger}.  \notag
\end{eqnarray}%
Obviously, the following relations should hold:
\begin{eqnarray}
\mathcal{P}H_{L,0}\mathcal{P}=H_{L,0} &\qquad \Longrightarrow \qquad &[%
\mathcal{P},H_{L,0}]=0
\nonumber
\\
\mathcal{P}H_{L,1}\mathcal{P}=-H_{L,1} &\qquad \Longrightarrow \qquad
&\left\{ \mathcal{P},H_{L,1}\right\} =0.  \label{a6}
\end{eqnarray}%
The first of these conditions together with $\mathcal{P}^{2}=I$, $\mathcal{P}%
\neq I$ reduces the possible form of the parity transformation to one of the
three types,
\begin{equation}
\mathcal{P}_{0x}:=I\otimes \sigma _{x}\,,\qquad \mathcal{P}_{x0}:=\sigma
_{x}\otimes I\,,\qquad \mathcal{P}_{xx}:=\sigma _{x}\otimes \sigma _{x}\,,
\label{a7}
\end{equation}%
where $\sigma _{x,y,z}$ are the usual Pauli matrices. Taking into
account that $\sigma _{x}\sigma _{z}\sigma _{x}=-\sigma _{z}$, the
anti-commutativity condition in Eqs. (\ref{a6}) singles out the only
possible parity matrix:
\begin{equation}
\mathcal{P}=\mathcal{P}_{x0}=
\left(\begin{array}{cc}
0 & I \\
I & 0
\end{array}%
\right)=
\left(
\begin{array}{cccc}
0 & 0 & 1 & 0 \\
0 & 0 & 0 & 1 \\
1 & 0 & 0 & 0 \\
0 & 1 & 0 & 0
\end{array}%
\right)
\label{a8}
\end{equation}%
for the linear transformation, i.e., the matrix interchanging $A$ and $C$,
as well as $B$ and $D$.
%\begin{eqnarray}
%u_{A} &\quad \mapsto \quad &\mathcal{P}u_{A}=u_{C}  \notag  \label{a9} \\
%u_{B} &\quad \mapsto \quad &\mathcal{P}u_{B}=u_{D}  \notag \\
%u_{C} &\quad \mapsto \quad &\mathcal{P}u_{C}=u_{A}  \notag \\
%u_{D} &\quad \mapsto \quad &\mathcal{P}u_{D}=u_{B}.
%\end{eqnarray}%
One then immediately checks that
\begin{equation}
 H_{NL}(\mathcal{P}\mathbf{u})=\mathcal{P}H_{NL}(\mathbf{u})\mathcal{P}%
=-\left(
\begin{array}{cccc}
|u_{C}|^{2} & 0 & 0 & 0 \\
0 & |u_{D}|^{2} & 0 & 0 \\
0 & 0 & |u_{A}|^{2} & 0 \\
0 & 0 & 0 & |u_{B}|^{2}%
\end{array}%
\right) \neq H_{NL}^{\dagger}(\mathbf{u}).\label{nl-2}
\end{equation}%
Hence, in contrast to the linear component $H_{L}$, the nonlinear terms $%
H_{NL}(\mathbf{u})$ corresponding to Eqs. (\ref{zpzm1}) are not $\mathcal{P}%
\mathcal{T}-$symmetric, in the usual matrix sense. Rather, the symmetry
properties of the nonlinear terms have to be considered in the context of
the nonlinear Schr\"{o}dinger equation itself. Acting with $\cP\bT$ on Eq. \rf{a1} we observe that
\begin{eqnarray}
\mathcal{P}\mathbf{T}(i\partial _{t}\mathbf{u}) &=&\mathcal{P}\mathbf{T}%
\left[ H_{L}\mathbf{u}+H_{NL}(\mathbf{u})\mathbf{u}\right] ,  \notag \\
i\partial _{t}(\mathcal{P}\mathbf{T}\mathbf{u}) &=&H_{L}\mathcal{P}\mathbf{T}%
\mathbf{u}+H_{NL}(\mathcal{P}\mathbf{T}\mathbf{u})(\mathcal{P}\mathbf{T}%
\mathbf{u})\label{a13}.
\end{eqnarray}
Hence, the full NLSE system (\ref{zpzm1}) remains invariant if we define the $\mathcal{P}\mathbf{T}$ transformation of the vectorial wave function obeying
this system as follows:
\begin{eqnarray}
\mathcal{P}\mathbf{T}\mathbf{u} &=&e^{i\phi }\mathbf{u},\qquad \phi \in
\mathbb{R}  \notag  \label{a13-2} \\
\mathcal{P}\bar{\mathbf{u}}(-t) &=&e^{i\phi }\mathbf{u}(t).
\end{eqnarray}
This is in full analogy to the condition of exact $\mathcal{P}\mathbf{T}%
- $symmetry for the corresponding linear Schr\"{o}dinger equation\footnote{We note that apart from trivial stationary type solutions with factorizing structure $\bu(t)=e^{-iEt}\bu_0$, nonlinearity matrices $H_{NL}(\bu)$ of more general type than that in \rf{nl-1} and \rf{nl-2} may be envisioned which may produce $\cP\bT-$symmetric solutions $\bu(t)$ with less simple time dependence. A detailed analysis of such systems will be presented elsewhere.}. But the
condition of \textit{spontaneously broken} $\mathcal{P}\mathbf{T}-$%
symmetry ($[\mathcal{P}\mathbf{T},H_{L}]=0$,\ $\mathcal{P}\mathbf{T}\mathbf{u%
}\not\propto \mathbf{u}$) is replaced by the condition of  \textit{%
completely broken} $\mathcal{P}\mathbf{T}-$symmetry ($\mathcal{P}\mathbf{T}%
\mathbf{u}\not\propto \mathbf{u}$\quad $\Longrightarrow $\quad $\lbrack
\mathcal{P}\mathbf{T},H_{NL}(\mathbf{u})]\neq 0$). In contrast to the
present 2D plaquettes, which are mainly motivated by feasible experimental
realizations, one can envision more sophisticated setups with \ $\mathcal{P}%
H_{NL}(\mathbf{u})\mathcal{P}=H_{NL}^{\dagger}(\mathbf{u})$. This will lead
to a new type of partial (or intermediate) $\mathcal{P}\mathbf{T}-$symmetry
(to be considered elsewhere), which for solutions $\mathbf{u}(t)$ with
broken $\mathcal{P}\mathbf{T}-$symmetry will keep the nonlinear term $H_{NL}(%
\mathbf{u})$ explicitly $\mathcal{P}\mathcal{T}-$symmetric ($\mathcal{P}-$%
pseudo-Hermitian) in the matrix sense, but not $\mathcal{P}\mathbf{T}-$%
symmetric (under inclusion of the explicit time reversal $t\rightarrow -t$)
in the sense of the NLSE system.

%This is in full analogy to the condition of exact $\cP\bT-$symmetry for the corresponding linear Schr\"odinger equation. But the condition of spontaneously broken $\cP\bT-$symmetry ($[\cP\bT,H_L]=0$,\ $\cP\bT\bu\not\propto \bu$) is replaced by the condition of completely broken $\cP\bT-$symmetry  ($\cP\bT\bu\not\propto \bu$\quad$\Longrightarrow$\quad $[\cP\bT,H_{NL}(\bu)]\neq 0$). In contrast to the present 2D plaquettes, which are mainly motivated by  possible direct experimental realizations, one can envision more sophisticated setups with $\cP H_{NL}(\bu)\cP=  H_{NL}^\dd(\bu)$. This will ensure a robust $\cP\bT-$symmetry so that for them the usual concepts of exact and spontaneously broken $\cP\bT-$symmetry extend straight forwardly to the NLSE as well\footnote{Corresponding studies will be the subject of  a separate work.}.

For configurations (b) and (c) in Fig.~\ref{modes}, $H_{L,0}$ and $H_{NL}(%
\mathbf{u})$ are still given by Eqs.~(\ref{a3}) and (\ref{a5}), but with
\begin{eqnarray}
H_{L,1} &=&\left(
\begin{array}{cccc}
i\gamma & 0 & 0 & 0 \\
0 & -i\gamma & 0 & 0 \\
0 & 0 & i\gamma & 0 \\
0 & 0 & 0 & -i\gamma
\end{array}%
\right) =i\gamma I\otimes \sigma _{z},  \label{a20} \\
H_{L,1} &=&\left(
\begin{array}{cccc}
i\gamma & 0 & 0 & 0 \\
0 & i\gamma & 0 & 0 \\
0 & 0 & -i\gamma & 0 \\
0 & 0 & 0 & -i\gamma
\end{array}%
\right) =i\gamma \sigma _{z}\otimes I
\end{eqnarray}%
respectively. Hence, relations (\ref{a6}) are valid for both configurations
(b) and (c) as well. Using (\ref{a7}) in $\mathcal{P}H_{L,1}\mathcal{P}%
=-H_{L,1}$, we find a richer variety of parity operators $\mathcal{P}$ than
for configuration (a). Configuration (b) allows for
\begin{equation}
\mathcal{P}_{0x}=\left(
\begin{array}{cc}
\sigma _{x} & 0 \\
0 & \sigma _{x}
\end{array}%
\right) ,\qquad \mathcal{P}_{xx}=\left(
\begin{array}{cc}
0 & \sigma _{x} \\
\sigma _{x} & 0
\end{array}%
\right) ,  \label{a21}
\end{equation}%
whereas configuration (c) may be associated with
\begin{equation}
\mathcal{P}_{x0}=\left(
\begin{array}{cc}
0 & I \\
I & 0
\end{array}%
\right) ,\qquad \mathcal{P}_{xx}=\left(
\begin{array}{cc}
0 & \sigma _{x} \\
\sigma _{x} & 0
\end{array}%
\right) .  \label{a21-2}
\end{equation}%
For configuration (d), we have
\begin{eqnarray}
H_{L,0} &=&\left(
\begin{array}{ccccc}
0 & 0 & -k & 0 & 0 \\
0 & 0 & -k & 0 & 0 \\
-k & -k & 0 & -k & -k \\
0 & 0 & -k & 0 & 0 \\
0 & 0 & -k & 0 & 0%
\end{array}%
\right) ,  \label{22} \\
H_{L,1} &=&\left(
\begin{array}{ccccc}
-i\gamma & 0 & 0 & 0 & 0 \\
0 & i\gamma & 0 & 0 & 0 \\
0 & 0 & 0 & 0 & 0 \\
0 & 0 & 0 & -i\gamma & 0 \\
0 & 0 & 0 & 0 & i\gamma%
\end{array}%
\right) , \\
H_{NL}(\mathbf{u}) &=&-\left(
\begin{array}{ccccc}
|u_{A}|^{2} & 0 & 0 & 0 & 0 \\
0 & |u_{B}|^{2} & 0 & 0 & 0 \\
0 & 0 & |u_{C}|^{2} & 0 & 0 \\
0 & 0 & 0 & |u_{D}|^{2} & 0 \\
0 & 0 & 0 & 0 & |u_{E}|^{2}
\end{array}%
\right) ,
\end{eqnarray}%
and simple computer algebra gives again two possible parity operators:
\begin{equation}
\mathcal{P}_{d,0}=\left(
\begin{array}{ccccc}
0 & 1 & 0 & 0 & 0 \\
1 & 0 & 0 & 0 & 0 \\
0 & 0 & 1 & 0 & 0 \\
0 & 0 & 0 & 0 & 1 \\
0 & 0 & 0 & 1 & 0%
\end{array}%
\right) ,\quad \mathcal{P}_{d,x}=\left(
\begin{array}{ccccc}
0 & 0 & 0 & 0 & 1 \\
0 & 0 & 0 & 1 & 0 \\
0 & 0 & 1 & 0 & 0 \\
0 & 1 & 0 & 0 & 0 \\
1 & 0 & 0 & 0 & 0%
\end{array}%
\right) ,
\end{equation}%
in strong structural analogy to configuration (b). One verifies that $H_{NL}(%
\mathcal{P}\mathbf{T}\mathbf{u})=\mathcal{P}H_{NL}(\mathbf{u})\mathcal{P}%
\neq H_{NL}^{\dagger}(\mathbf{u})$ holds also for configurations (b), (c)
and (d), hence all 2D plaquettes considered in the present paper are not $%
\mathcal{P}\mathbf{T}-$symmetric in the usual matrix sense.
%, but only in the NLSE sense of relations \rf{a13}, \rf{a13-2}.

\subsection{Spectral behavior of associated linear setups}

Next, we turn to the eigenvalue problems of the linear setups associated
with plaquettes (a) - (d), i.e., to solutions of the equation
\begin{equation}
H_{L}\mathbf{u}_{n}=E_{n}\mathbf{u}_{n}.  \label{a14}
\end{equation}%
From the corresponding characteristic polynomials, $\det (H_{L}-EI)=0$,
\begin{align}
(\mathrm{a}):\qquad \qquad & E^{2}\left[ E^{2}-(4k^{2}-\gamma ^{2})\right]
=0,  \notag \\
(\mathrm{b}):\qquad \qquad & (E^{2}+\gamma ^{2})\left[ E^{2}-(4k^{2}-\gamma
^{2})\right] =0,  \notag \\
(\mathrm{c}):\qquad \qquad & E^{4}-2(2k^{2}-\gamma ^{2})E^{2}+\gamma ^{4}=0,
\notag \\
(\mathrm{d}):\qquad \qquad & E(E^{2}+\gamma ^{2})\left[ E^{2}-(4k^{2}-\gamma
^{2})\right] =0,
\end{align}%
we find
\begin{eqnarray}
(\mathrm{a}):\qquad \qquad &&E_{1,2}=0,\qquad E_{3,4}=\pm \sqrt{%
4k^{2}-\gamma ^{2}}  \notag  \label{a15-2} \\
(\mathrm{b}):\qquad \qquad &&E_{1,2}=\pm i\gamma ,\qquad E_{3,4}=\pm
\sqrt{4k^{2}-\gamma ^{2}}  \notag \\
(\mathrm{c}):\qquad \qquad &&E_{1,2}=\sqrt{2k^{2}-\gamma ^{2}\pm 2k\sqrt{%
k^{2}-\gamma ^{2}}}  \notag \\
&&E_{3,4}=-\sqrt{2k^{2}-\gamma ^{2}\pm 2k\sqrt{k^{2}-\gamma ^{2}}}  \notag \\
(\mathrm{d}):\qquad \qquad &&E_{1,2}=\pm i\gamma ,\qquad E_{3,4}=\pm
\sqrt{4k^{2}-\gamma ^{2}},\qquad E_{5}=0.
\end{eqnarray}%
Obviously, the matrix Hamiltonians $H_{L}$ for plaquettes (a) and (d) are not of
full rank. For plaquette (a) we find $\mbox{\rm rank}\,(H_{L})=2$, and $%
H_{L}$ has a two-dimensional kernel space, $\ker (H_{L})=\,\mbox{\rm span}_{%
\mathbb{C}}(\mathbf{u}_{1},\mathbf{u}_{2})$. For plaquette (d) we find $%
\mbox{\rm
rank}\,(H_{L})=4$ and $\ker (H_{L})=\mathbb{C}^{\ast }\times \mathbf{u}_{5}$%
, where $\mathbb{C}^{\ast }=\mathbb{C}-\{0\}$. Moreover, we see that the
spectrum for plaquette (d), up to the additional eigenvalue $E_{5}=0$,
coincides with that for (b).
The different eigenvalues of the 4-node plaquettes  displayed in Eqs. \rf{a15-2} show that these plaquettes are also physically not equivalent. Equivalence classes of nonlinear 4-node plaquettes with isospectral linear Hamiltonians $H_L$ but different pairwise couplings have been considered,  e.g., in~\cite{konorecent4}.
For plaquettes (a), (b) and (d) an exceptional
point (EP) occurs at $\gamma ^{2}=4k^{2}$, being associated with a branching
of the eigenvalue pair $E_{3,4}$
\begin{eqnarray}
E_{3,4}\in \mathbb{R} &\qquad \mbox{\rm for}\qquad &4k^{2}\geq \gamma ^{2}
\notag  \label{a15-3} \\
E_{3,4}\in i\mathbb{R} &\qquad \mbox{\rm for}\qquad &4k^{2}<\gamma ^{2}.
\end{eqnarray}%
In the case of plaquette (a), all four eigenvalues are involved in the
branching at $\gamma =\pm 2k$, where $E_{1}=\ldots =E_{4}=0$. Via Jordan
decomposition (e.g., with the help of the corresponding linear algebra tool
of Mathematica) we find that
\begin{equation}
(a):\qquad \qquad H_{L}(\gamma =\pm 2k)\sim \left(
\begin{array}{cccc}
0 & 1 & 0 & 0 \\
0 & 0 & 1 & 0 \\
0 & 0 & 0 & 0 \\
0 & 0 & 0 & 0%
\end{array}%
\right) =J_{3}(0)\oplus J_{1}(0),  \label{a16}
\end{equation}%
i.e., a spectral degeneration of the type $(0^{3},0^{1})$ in  Arnold's
notation \cite{Arnold-degen}, or, in other words, a third-order EP with a
single decoupled fourth mode. Hence, plaquettes of type (a) may serve as an easily implementable testground for the experimental investigation of third-order EPs (see e.g. \cite{GS2005,heissEP3,GGKN2008,grae4}). For plaquettes (b) and (d) we have
second-order EPs at $\gamma ^{2}=4k^{2}$, similar as for plaquette (c) where
a pair of second-order EPs occurs at $\gamma ^{2}=k^{2}$ with $%
E_{1}=E_{2}=|k|$, $E_{3}=E_{4}=-|k|$.

From the eigenvalues in (\ref{a15-2}) we read off the $\mathcal{P}\mathbf{T}%
- $symmetry content of the four types of plaquettes. The sector of exact $%
\mathcal{P}\mathbf{T}-$symmetry (i.e., the sector with all eigenvalues
purely real, $E_{n}\in \mathbb{R},\ \forall n$) corresponds to
\begin{eqnarray}
(\mathrm{a}):\qquad \qquad &&\gamma ^{2}\leq 4k^{2},  \notag  \label{a16a} \\
(\mathrm{b}):\qquad \qquad &&\gamma =0,  \notag \\
(\mathrm{c}):\qquad \qquad &&\gamma ^{2}\leq k^{2},  \notag \\
(\mathrm{d}):\qquad \qquad &&\gamma =0,
\end{eqnarray}%
i.e., for plaquettes (b) and (d) the $\mathcal{P}\mathbf{T}-$symmetry is
spontaneously broken as soon as the gain-loss coupling is switched on, namely
for $%
\gamma \neq 0$.

\section{Existence, stability and dynamics of nonlinear states\label%
{numerics}}

In this section, we seek stationary solutions of the type
\begin{equation}
\mathbf{u}_{0}(t)=e^{-iEt}\mathbf{u}_{0},\qquad E\in \mathbb{R},\qquad
\mathbf{u}_{0}=(a,b,c,d)^{T}\in \mathbb{C}^{4}  \label{E}
\end{equation}%
constructed over constant vectors $\mathbf{u}_{0}$. According to \rf{a13} and \rf{a13-2} such solutions will be $\cP\bT-$symmetric provided it holds $\cP\cT\bu_0=e^{i\varphi}\bu_0$ for some $\varphi\in\RR$. We will test these symmetry properties for the solutions to be obtained. 
We note that restricting the explicit analysis to stationary solutions of the type \rf{E} we by construction exclude from this analysis $\cP\bT-$violating solutions with $E\not\in\RR$ which are necessarily non-stationary.

A useful technical tool to facilitate the explicit derivation of stationary solutions $\mathbf{u}_{0}(t)$ are conservation equations of the type
\begin{equation}
\partial _{t}(\mathbf{u}^{\dagger}Y\mathbf{u})=i\mathbf{u}^{\dagger}\left(
H^{\dagger}Y-YH\right) \mathbf{u,}
\end{equation}%
constructed from Eq. (\ref{a1}) and its adjoint, where $Y$ denotes an
arbitrary constant matrix. The most simplest of them can be found via Eqs. (\ref{a2-4}), (\ref{a3}) and
Eq. (\ref{a5}) to be
\begin{eqnarray}  \label{a18}
\partial _{t}|\mathbf{u}|^{2}=\partial _{t}(\mathbf{u}^{\dagger}\mathbf{u})
&=&-2i\mathbf{u}^{\dagger}H_{L,1}\mathbf{u}  \label{a18-1} \\
\partial _{t}(\mathbf{u}^{\dagger}\mathcal{P}\mathbf{u}) &=&i\mathbf{u}%
^{\dagger}\left[ H_{NL}^{\dagger}(\mathbf{u})\mathcal{P}-\mathcal{P}H_{NL}(%
\mathbf{u})\right] \mathbf{u}.  \label{a18-2}
\end{eqnarray}%
For stationary equations $\bu_0(t)$ the time-dependent phase factors $e^{-iEt}$ cancel so that the left-hand-sides of these relations vanish, yielding simple algebraic constraints on the right-hand-sides. From Eq. \rf{a18-2}  we see that for stationary solutions the $\mathcal{P}\mathcal{T}$ inner
product\footnote{For completeness, we note that in the context of $\cP\cT$ quantum mechanics (PTQM) the $\cP\cT$ inner product was introduced first by Znojil in \cite{znojil-2001} in 2001. Immediately afterwards, it was interpreted by Japaridze as indefinite inner product \cite{japa-2001} in a Krein space and generalized by Mostafazadeh to the $\eta-$metric in the context of pseudo-Hermitian Hamiltonians \cite{ali-2001}. Finally, it was used by Bender, Brody and Jones in 2002 to construct the positive definite $\cC\cP\cT$ inner product \cite{cmb-2002}. For oligomer settings (of plaquettes or other few site configurations), it can be employed, e.g.,  to derive a simple algebraic constraint or as a criterion of the numerical accuracy of the evolutionary dynamics  (especially since the solutions rapidly acquire very
large amplitudes when unstable, as will be seen below). It also turned out useful in \cite{konorecent4}.} will remain conserved ($\mathbf{u}^{\dagger}\mathcal{P}\mathbf{u}=$%
const) regardless of the violated $\cP-$pseudo-Hermiticity,   \ $%
\mathcal{P}H_{NL}^{\dagger}(\mathbf{u})\mathcal{P}\neq H_{NL}(\mathbf{u})$, characteristic for of our specific nonlinear plaquette couplings (see Eq. \rf{nl-2}).

Subsequently, we first derive classes of 
stationary solutions $\mathbf{u}_{0}(t)$ explicitly. Then, we analyze the
stability of small perturbations over these stationary solutions by the
linearization, via ansatz
\begin{equation}
\mathbf{u}(t)=e^{-iEt}\left[ \mathbf{u}_{0}+\delta (e^{\lambda t}r+e^{\bar{%
\lambda}t}s)\right] +O(\delta ^{2}),\qquad |\delta |\ll 1,
\end{equation}%
where $\delta $ is the small amplitude of the perturbation. Exponents $%
\lambda $ can be defined as Wick-rotated eigenvalues from the corresponding $%
8\times 8$ perturbation matrix $\mathbf{B}$ (see, e.g., \cite{pgk_book} for
more details):

\begin{eqnarray}
(\mathbf{B}-i\lambda I_{8})\mathbf{x} &=&0  \notag  \label{a24} \\
\mathbf{B}:= &&\left(
\begin{array}{cc}
\partial _{u_{n}}F(\mathbf{u},\bar{\mathbf{u}}) & \partial _{\bar{u}_{n}}F(%
\mathbf{u},\bar{\mathbf{u}}) \\
-\partial _{u_{n}}\bar{F}(\mathbf{u},\bar{\mathbf{u}}) & -\partial _{\bar{u}%
_{n}}\bar{F}(\mathbf{u},\bar{\mathbf{u}})%
\end{array}%
\right) ,\qquad n=1,2,3,4  \notag \\
\mathbf{x} &=&(r,\bar{s})^{T},
\end{eqnarray}%
where
\begin{equation}
F(\mathbf{u},\bar{\mathbf{u}}):=[H(\mathbf{u})-E]\mathbf{u}
\end{equation}%
characterizes the stationary problem, and the elements of the matrix $%
\mathbf{B}$ are evaluated at $\mathbf{u}=\mathbf{u}_{0}$. Linear stability
is ensured for $\lambda \in i\mathbb{R}$, whereas $\lambda \not\in i\mathbb{R%
}$ corresponds to growing and decaying modes, i.e., exponential
instabilities.

\subsection{The plaquette of the 0+0- type}

Substituting ansatz (\ref{E}) for the stationary solutions in Eqs.
(\ref{a1}), (\ref{a18-1}) and (\ref{a18-2}) we obtain the following
algebraic equations:
\begin{eqnarray}
Ea &=&k(b+d)+|a|^{2}a,  \notag \\
Eb &=&k(a+c)+|b|^{2}b-i\gamma b,  \notag \\
Ec &=&k(b+d)+|c|^{2}c,  \notag \\
Ed &=&k(a+c)+|d|^{2}d+i\gamma d,  \label{zpzm2}
\end{eqnarray}%
\begin{eqnarray}
\partial _{t}|\mathbf{u}|^{2} &=&-2i\mathbf{u}^{\dagger}H_{L,1}\mathbf{u,}
\notag  \label{a30} \\
0 &=&2\gamma (|b|^{2}-|d|^{2}),
\end{eqnarray}%
and
\begin{eqnarray}
\partial _{t}(\mathbf{u}^{\dagger}\mathcal{P}\mathbf{u}) &=&i\mathbf{u}%
^{\dagger}\left[ H_{NL}^{\dagger}(\mathbf{u})\mathcal{P}-\mathcal{P}H_{NL}(%
\mathbf{u})\right] \mathbf{u,}  \notag  \label{a31} \\
0 &=&\left( |a|^{2}-|c|^{2}\right) (\bar{a}c-\bar{c}a)+\left(
|b|^{2}-|d|^{2}\right) (\bar{b}d-\bar{d}b).
\end{eqnarray}%
These equations can be analyzed via the Madelung substitution (i.e., via
amplitude-phase decomposition),
\begin{equation}
a=Ae^{i\phi _{a}},b=Be^{i\phi _{b}},c=Ce^{i\phi _{c}},d=De^{i\phi _{d}}.
\label{aA}
\end{equation}%
Without loss of generality, we may fix $\phi _{a}=0$.

For arbitrary phase factors in (\ref{aA}), Eqs. (\ref{a30}) and (\ref{a31})
are satisfied by $A=C$ and $B=D$. Using this condition in Eq. (\ref{zpzm2})
and dividing each equation (\ref{zpzm2}) by the phase factor on its
left-hand side, one obtains the imaginary parts of the resulting equations:
\begin{eqnarray}
0 &=&kB\left[ \sin (\phi _{b}-\phi _{a})+\sin (\phi _{d}-\phi
_{a})\right] =2kB\sin \left( \frac{\phi _{b}+\phi _{d}}{2}-\phi
_{a}\right) \cos \left(
\frac{\phi _{b}-\phi _{d}}{2}\right) ,  \notag \\
\gamma B &=&kA\left[ \sin (\phi _{a}-\phi _{b})+\sin (\phi _{c}-\phi _{b})%
\right] =2kA\sin \left( \frac{\phi _{a}+\phi _{c}}{2}-\phi _{b}\right) \cos
\left( \frac{\phi _{a}-\phi _{c}}{2}\right) ,  \notag \\
0 &=&kB\left[ \sin (\phi _{b}-\phi _{c})+\sin (\phi _{d}-\phi
_{c})\right] =2kB\sin \left( \frac{\phi _{b}+\phi _{d}}{2}-\phi
_{c}\right) \cos \left(
\frac{\phi _{b}-\phi _{d}}{2}\right) ,  \notag \\
-\gamma B &=&kA\left[ \sin (\phi _{a}-\phi _{d})+\sin (\phi
_{c}-\phi _{d})\right] =2kA\sin \left( \frac{\phi _{a}+\phi
_{c}}{2}-\phi _{d}\right)
\cos \left( \frac{\phi _{a}-\phi _{c}}{2}\right) .  \notag \\
&&  \label{zpzm_add}
\end{eqnarray}%
For $\phi _{a}=0$ the first of these equations implies $\sin (\phi
_{b})=-\sin (\phi _{d})$, hence either $\phi _{b}=-\phi _{d}$ (case 1) or $%
\phi _{d}=\phi _{b}-\pi $ (case 2). In case 1, we conclude from the third
equation that either $\phi _{b}\neq \pm \pi /2$ and $\phi _{c}=0$ (case 1a),
or $\phi _{b}=\pm \pi /2$ and $\phi _{c}$ is arbitrary (case 1b). In case 2
the third equation is satisfied automatically. In all the three cases, the
second and the fourth equation are compatible. They give
\begin{eqnarray}
\mbox{case 1a:}\qquad \qquad &&\sin (\phi _{b})=-\frac{\gamma B}{2kA}%
,\qquad \phi _{c}=0,\qquad \phi _{d}=-\phi _{b},  \notag  \label{a32} \\
\mbox{case 1b:}\qquad \qquad &&\cos (\phi _{c})=\mp \frac{\gamma B}{kA}%
-1,\qquad \phi _{d}=-\phi _{b}=\mp \pi /2,  \notag \\
\mbox{case 2:}\qquad \qquad &&\sin (\phi _{b})+\sin (\phi _{b}-\phi
_{c})=-\frac{\gamma B}{kA},\qquad \phi _{d}=\phi _{b}-\pi .
\end{eqnarray}%
Returning to the phase-factor divided equations (\ref{zpzm2}) and
considering their real parts, we find
\begin{eqnarray}
EA &=&kB\left[ \cos (\phi _{b}-\phi _{a})+\cos (\phi _{d}-\phi _{a})\right]
+A^{3},  \notag  \label{a33} \\
EB &=&kA\left[ \cos (\phi _{a}-\phi _{b})+\cos (\phi _{c}-\phi _{b})\right]
+B^{3},  \notag \\
EA &=&kB\left[ \cos (\phi _{b}-\phi _{c})+\cos (\phi _{d}-\phi _{c})\right]
+A^{3},  \notag \\
EB &=&kA\left[ \cos (\phi _{a}-\phi _{d})+\cos (\phi _{c}-\phi _{d})\right]
+B^{3}.
\end{eqnarray}%
The pairwise compatibility of the first and third, as well as of the second
and fourth equations requires
\begin{eqnarray}
\cos (\phi _{b}-\phi _{a})+\cos (\phi _{d}-\phi _{a}) &=&\cos (\phi
_{b}-\phi _{c})+\cos (\phi _{d}-\phi _{c}),  \notag  \label{a34} \\
\cos (\phi _{a}-\phi _{b})+\cos (\phi _{c}-\phi _{b}) &=&\cos (\phi
_{a}-\phi _{d})+\cos (\phi _{c}-\phi _{d}).
\end{eqnarray}%
For case 1a, these conditions are trivially satisfied, whereas for the
remaining cases they lead to further restrictions:
\begin{eqnarray}
\mbox{case 1b:}\qquad &\phi _{c}=0;\pi \qquad \Longrightarrow
&\qquad
\gamma =\mp \frac{2kA}{B};  \notag  \label{a35} \\
 &&\qquad \gamma =0;  \notag \\
\mbox{case 2:} &&\quad \phi _{c}=2\phi _{b}\pm \pi ,\quad \sin (\phi
_{b})=-\frac{\gamma B}{2kA}.
\end{eqnarray}

In this way the phase angles are fixed for all the three cases and we can
turn to the amplitudes. The corresponding equation sets reduce to
\begin{eqnarray}
\mbox{case 1a:}\qquad \qquad EA &=&2kB\cos (\phi _{b})+A^{3},  \notag
\label{a36} \\
\qquad \qquad \qquad \quad EB &=&2kA\cos (\phi _{b})+B^{3},  \notag \\
\mbox{case 1b,2:}\qquad \qquad E &=&A^{2}=B^{2}.
\end{eqnarray}%
\ In the latter two cases (1b and 2) the amplitudes and phases completely
decouple and we have
\begin{equation}
A=B=C=D=\sqrt{|E|}.
\end{equation}%
Case 1a allows for a richer behavior. Equating the terms $2k\cos (\phi _{b})$
in the upper two equations (\ref{a36}) leads to the constraint
\begin{equation}
A^{2}(E-A^{2})=B^{2}(E-B^{2}),  \label{a38}
\end{equation}%
which can be resolved by $A=B$ (case 1aa) as well as by $E=A^{2}+B^{2}$
(case 1ab). The analysis of these two cases can be completed with the help
of the relation $\cos (\phi _{b})=\pm \sqrt{1-\frac{\gamma ^{2}B^{2}}{%
4k^{2}A^{2}}}$ from Eq. (\ref{a32}).

As result we obtain the following set of stationary solutions:
\begin{eqnarray}
\mbox{case 1a:}\qquad &&\sin (\phi _{b})=-\frac{\gamma
B}{2kA},\qquad
\phi _{c}=0,\qquad \phi _{d}=-\phi _{b},  \notag  \label{a39} \\
\mbox{case 1aa:}\qquad &&A=B=C=D=\sqrt{E\mp \sqrt{4k^{2}-\gamma
^{2}}},
\label{a39-1aa} \\
\mbox{case 1ab:}\qquad &&A=C,\ B=D=\frac{2kA}{\sqrt{A^{4}+\gamma ^{2}}}%
,\quad E=A^{2}+B^{2},  \label{a39-1ab} \\
\mbox{case 1b:}\qquad &&\phi _{d}=-\phi _{b}=\mp \pi /2,\quad \phi
_{c}=0,\pi ,\qquad \gamma =\pm 2k,\quad \gamma =0,  \notag \\
 &&A=B=C=D=\sqrt{E},  \label{a39-1b} \\
\mbox{case 2:}\qquad &&\sin (\phi _{b})=-\frac{\gamma }{2k},\qquad
\phi
_{d}=\phi _{b}-\pi ,\qquad \phi _{c}=2\phi _{b}\pm \pi ,  \notag \\
 &&A=B=C=D=\sqrt{E}.  \label{a39-2}
\end{eqnarray}

From Eq.~(\ref{a31}), it can also be seen that either $A=C$ or if
$A\neq C$, then $\sin (\phi _{a}-\phi _{c})=0$ must be true.
Here, we use the
information available so far to check the $\mathcal{P}\mathbf{T}-$symmetry
content of the solutions (\ref{a39-1aa}) - (\ref{a39-2}) explicitly. For
stationary solutions $\mathbf{u}(t)=e^{-iEt}\mathbf{u}_{0}$, $E\in \mathbb{R}$
the $\mathcal{P}\mathbf{T}-$symmetry condition (\ref{a13-2}) implies
\begin{equation}
\mathcal{P}\mathcal{T}\mathbf{u}_{0}=e^{i\phi }\mathbf{u}_{0}.
\end{equation}%
Taking into account that $\mathcal{T}$ acts as complex conjugation, we see
from the explicit structure of $\mathcal{P}=\mathcal{P}_{x0}$ in Eq. (\ref%
{a8})
%%%%%and (\ref{a9})
that a stationary solution is $\mathcal{P}\mathbf{T}-$%
symmetric if, with $\phi _{a}=0$, it has $\phi _{c}=0$ and $\phi _{d}=-\phi
_{b}$ (up to a common phase shift). Additionally, the amplitudes have to
coincide pairwise: $A=C$, $B=D$.
%For \rf{a39-1aa} - \rf{a39-2} this means that all case-1 stationary solutions with $\phi_c=0$ are $\cP\bT-$symmetric, whereas case 2 corresponds to a mode with broken $\cP\bT-$symmetry. The algebraically more involved modes with $A\neq C$ are also not $\cP\bT-$symmetric.
For Eqs. (\ref{a39-1aa}) - (\ref{a39-2}) this means that all case-1
stationary solutions with $\phi _{c}=0$ are $\mathcal{P}\mathbf{T}-$%
symmetric in their present form. The case-2 mode becomes explicitly $%
\mathcal{P}\mathcal{T}-$symmetric after a global $U(1)$ multiplication by a
phase factor:
\begin{eqnarray}
\mathbf{u}_{0} &=&e^{i(\phi _{b}-\pi /2)}\mathbf{v}_{0},  \notag  \label{a41}
\\
\mathbf{v}_{0}:= &&A\left[ e^{-i(\phi _{b}-\pi /2)},e^{i\pi /2},e^{i(\phi
_{b}-\pi /2)},e^{-i\pi /2}\right] ^{T},  \notag \\
\mathcal{P}\mathcal{T}\mathbf{v}_{0} &=&\mathbf{v}_{0},
\end{eqnarray}%
where $\phi _{c}=2\phi _{b}-\pi $ has to be chosen in Eq. (\ref{a39-2}). We
note that this procedure is effectively equivalent to a redefinition of the
original phase constraint: $\phi _{a}=0\ \mapsto \ \phi _{a}=-\phi _{b}+\pi
/2$ at the very beginning of the calculations in Eq. (\ref{aA}).

The linear stability analysis was performed numerically. Subsequently we
present corresponding graphical results. The plaquettes (b) - (d) can be
analyzed in a similar way. For brevity's sake, in Fig. \ref{figzpzm1} we
present only the basic numerical results, by means of the following symbols:

\begin{figure}[htp]
\label{figzpzm1}\scalebox{0.4}{\includegraphics{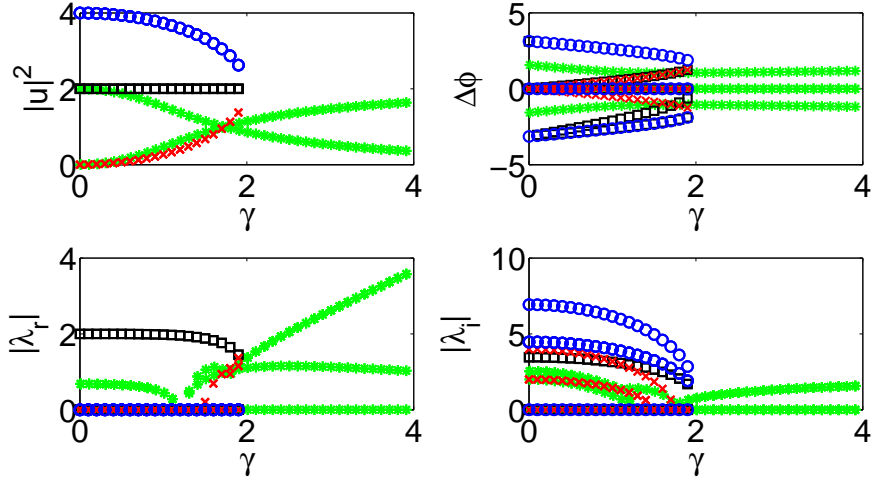}}
\caption{(Color online) Profiles of the solutions for plaquette (a) from Fig.
\protect\ref{modes}, with $E=2$ and $k=1$. Four different branches of the
solutions are denoted by blue circles, red crosses, black squares and green
stars. The top left and right panel display, respectively, the squared
absolute values of the amplitudes and phase differences between adjacent
sites for the respective states. The bottom left and right panels show real
(the instability growth rates) and imaginary (oscillation frequencies) parts
of the eigenvalues produced by the linearization around the stationary
states. The continuations are shown versus the gain-loss parameter $\protect%
\gamma $. }
%In the top right panel, the squared black is overlapped by the top
%crossed red curve and the bottom circled blue curve,
%and the middle crossed red line overlaps the stared green one.}
\end{figure}

\begin{itemize}
\item case 1aa with $A=B=C=D=\sqrt{E+\sqrt{4k^{2}+\gamma ^{2}}}$ \qquad ---
blue circles;

\item case 1aa with $A=B=C=D=\sqrt{E-\sqrt{4k^{2}-\gamma ^{2}}}$ \qquad ---
red crosses;

\item case 1ab --- green stars;

\item case 2 --- black squares;

\item Case 1b is not depicted explicitly because it corresponds to point
configurations without gain-loss ($\gamma =0$) and to exceptional point
configurations $\gamma =\pm 2k$.
\end{itemize}

\begin{figure}[tph]
\scalebox{0.4}{\includegraphics{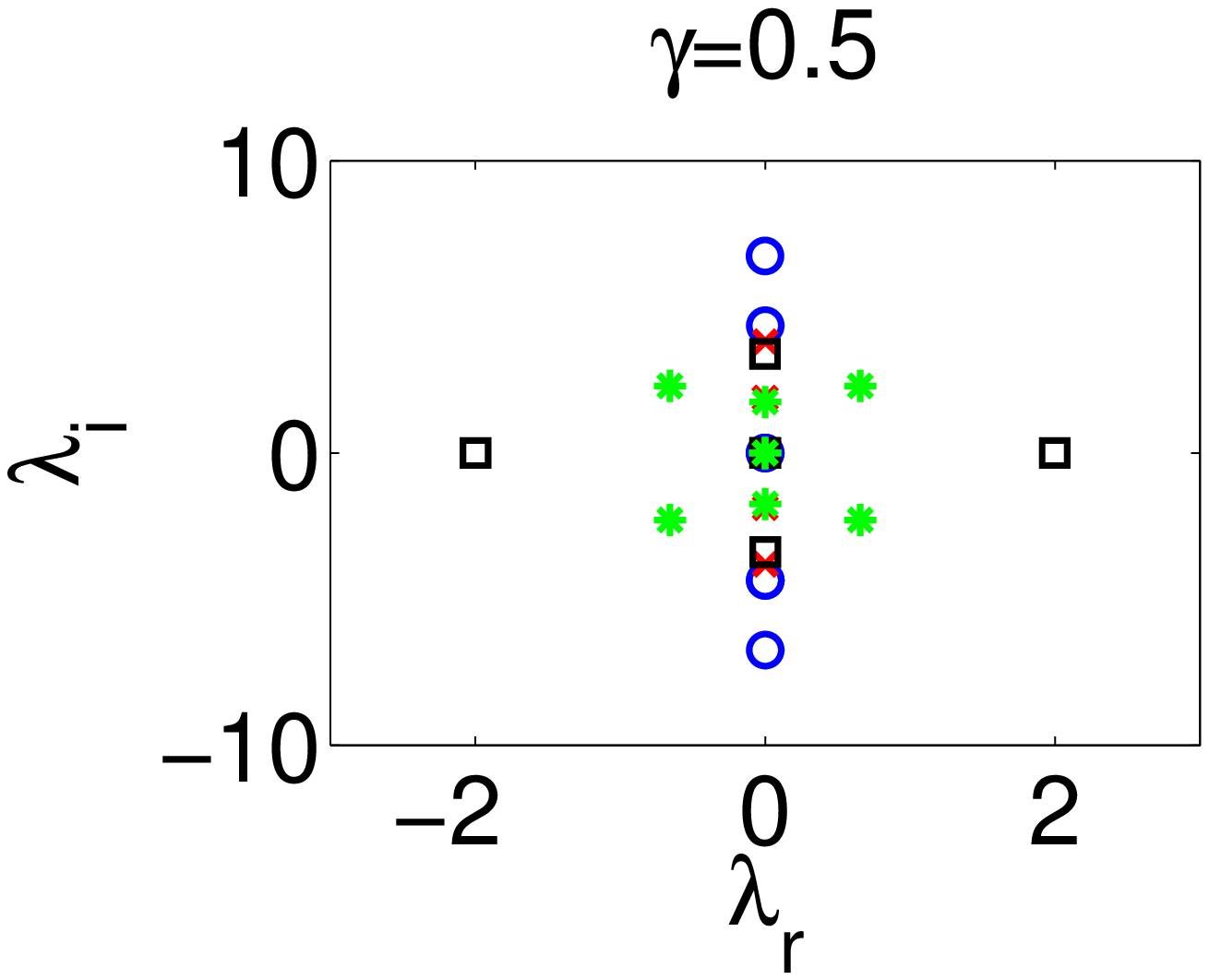}} \hfill\scalebox{0.4}{%
\includegraphics{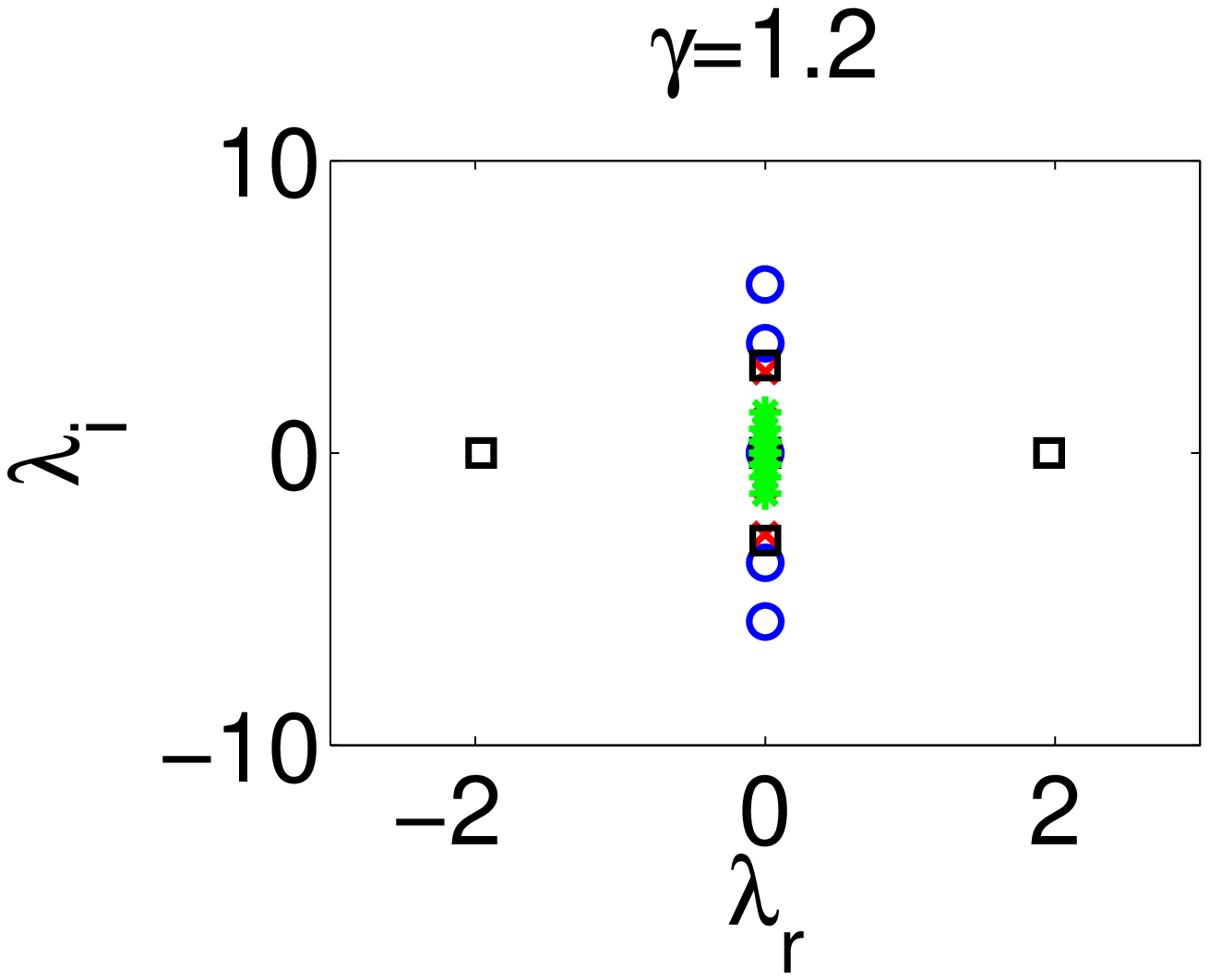}} \newline
\scalebox{0.4}{\includegraphics{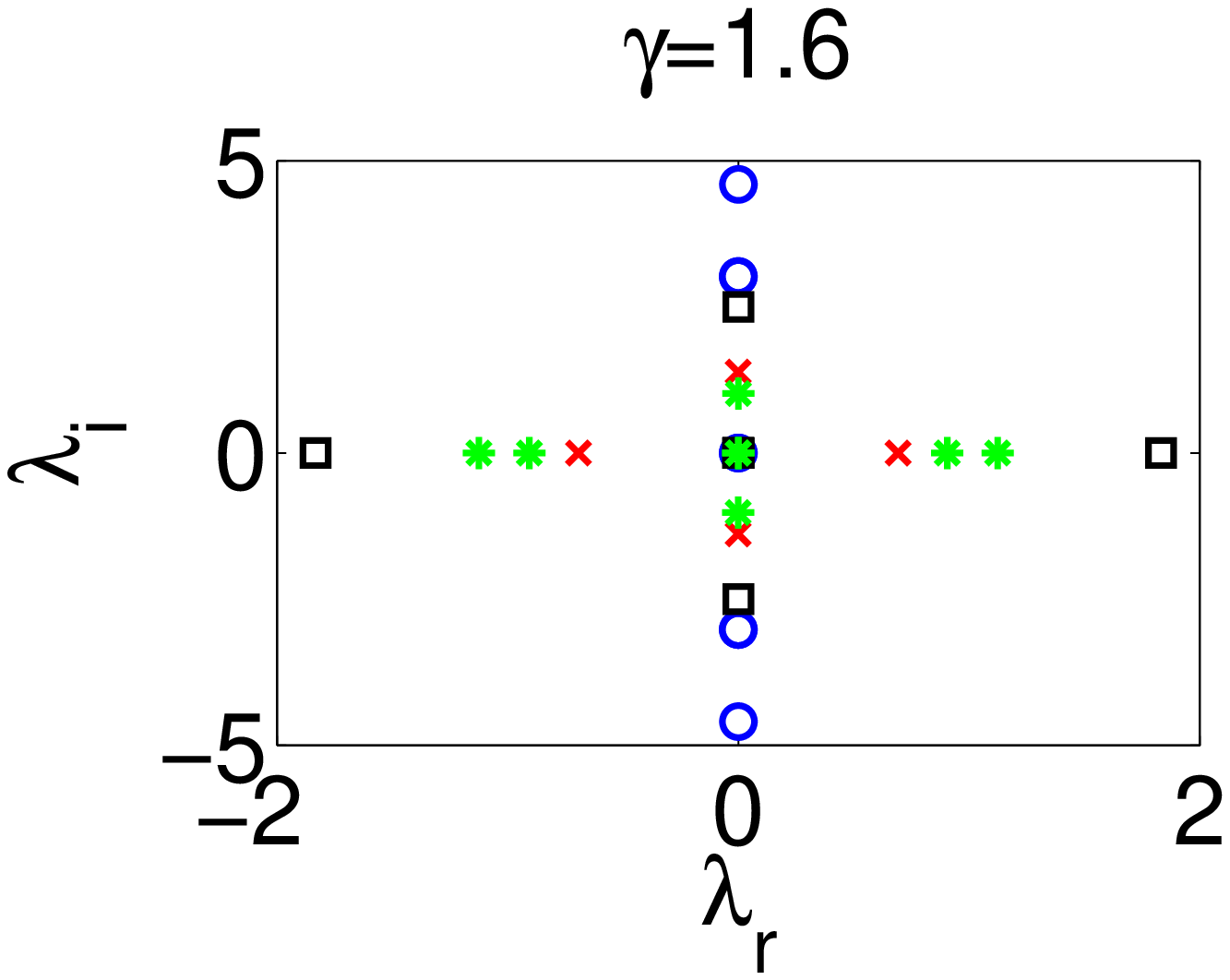}} \hfill \scalebox{0.4}{%
\includegraphics{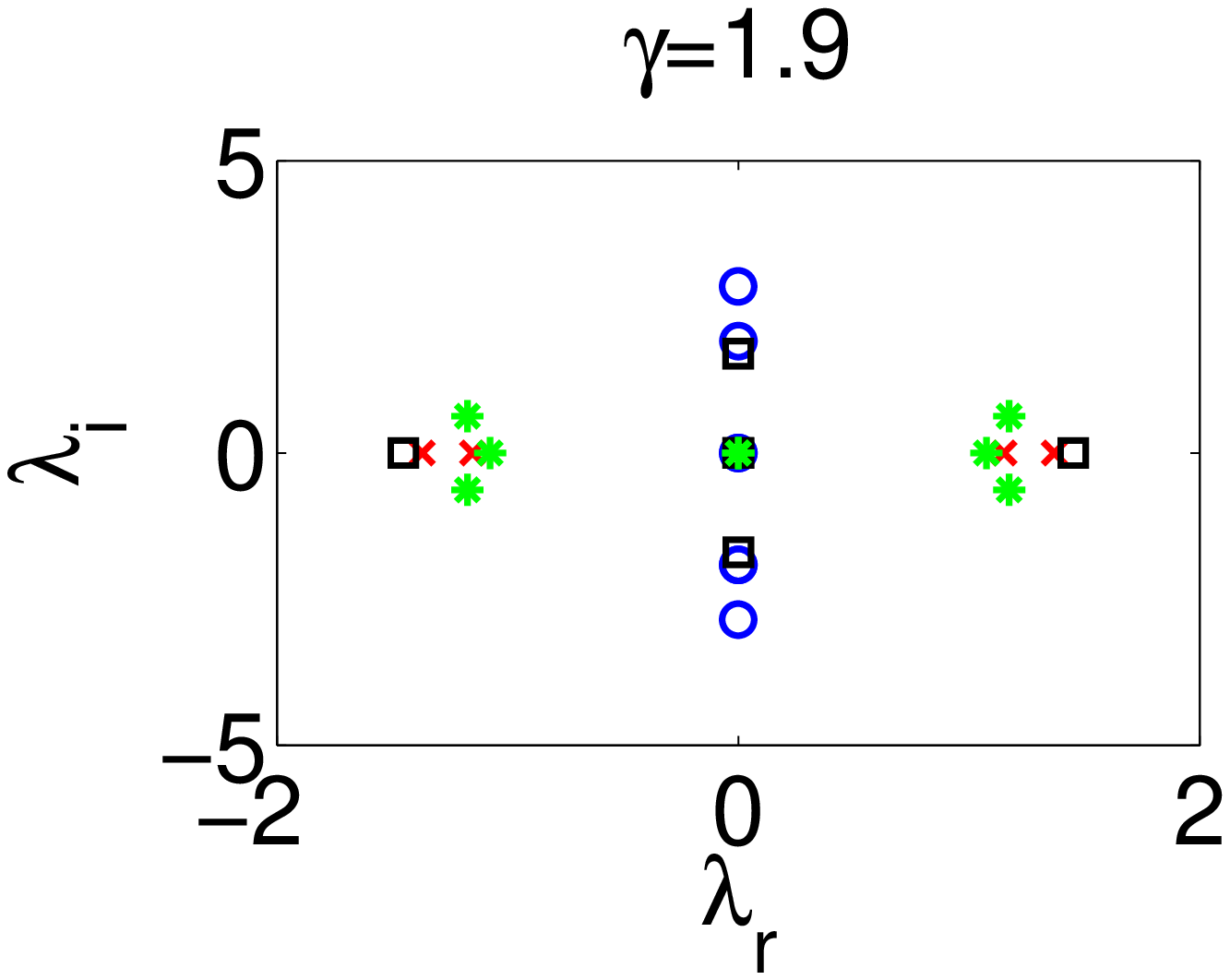}}
\caption{(Color online) The stability plots for plaquette (a) from Fig.
\protect\ref{modes} with $E=2$ and $k=1$, for different values of $\protect%
\gamma $. The notation for different branches is the same as in the previous
figure. All branches are shown for $\protect\gamma =0.5$, $\protect\gamma %
=1.2$, $\protect\gamma =1.6$, and $\protect\gamma =1.9$ (top left, top
right, bottom left, and bottom right panels, respectively).}
\label{figzpzm2}
\end{figure}

Figure~\ref{figzpzm1} presents the mode branches (their amplitudes, phases,
and also their stability) over the gain-loss parameter $\gamma $, starting
from the conservative system at $\gamma =0$. The same symbols are used in
Fig.~\ref{figzpzm2}, which displays typical examples of the spectral plane $%
(\lambda _{r},\lambda _{i})$ for stability eigenvalues $\lambda =\lambda
_{r}+i\lambda _{i}$ of the linearization; recall that the modes are unstable
if they give rise to $\lambda _{r}\neq 0$. Explicitly we observe the
following behavior.

\begin{itemize}
\item case 1aa with $A=B=C=D=\sqrt{E+\sqrt{4k^{2}+\gamma ^{2}}}$ \qquad ---
blue circles\newline
According to Fig. \ref{figzpzm2}, the present solution is stable. Notice
that, although featuring a phase profile, it cannot be characterized as a
vortex state (the same is true for some other configurations carrying
phase structure). Interestingly, the relevant configuration is generically
stable bearing two imaginary pairs of eigenvalues.
%  \item case 1aa with $A=B=C=D=\sqrt{E-\sqrt{4k^2-\g^2}}$ \qquad --- red crosses\\
%Obviously, this solution as well as the previous one exist up to the exceptional point $\g=\pm 2k$
%of the $\mathcal{PT}$-symmetry breaking in the linear system, where
%the two branches collide and disappear. As seen in Fig. \ref{figzpzm2}, the
%present branch has two eigenvalue pairs which are purely imaginary for small
%$\gamma $, but become real (rendering the configuration unstable),
%respectively, at $\gamma =1.49$ and then $\gamma =1.73$. Ultimately, these
%pairs of unstable eigenvalues collide at the origin of the spectral plane
%with those of the previous branch (blue circles).

\item case 1aa with $A=B=C=D=\sqrt{E-\sqrt{4k^{2}-\gamma ^{2}}}$ \qquad ---
red crosses.\newline
Obviously, this kind of solutions as well as the previous one exist up to the
exceptional point $\gamma =\pm 2k$ of the $\mathcal{PT}$-symmetry breaking
in the linear system, where the two branches collide and disappear (leave
the stationary regime and become nonstationary). As seen in Fig. \ref%
{figzpzm2}, the present branch has two eigenvalue pairs which are purely
imaginary for small $\gamma $, but become real (rendering the configuration
unstable) at $\gamma =1.49$ and then $\gamma =1.73$, respectively.
Ultimately, these pairs of unstable eigenvalues collide at the origin of the
spectral plane with those of the previous branch (blue circles).

\item case 1ab --- green stars.\newline
This stationary solution has a number of interesting features. Firstly, it
is the only one among the considered branches which has two unequal
amplitudes. Secondly, it exists past the critical point $\gamma =\pm 2k$ of
the linear system, due to the effect of the nonlinearity (the extension of
the existence region for nonlinear modes was earlier found in 1D couplers
\cite{miron} and oligomers~\cite{pgk,konorecent4}). Furthermore, this branch
has three non-zero pairs of stability eigenvalues, two of which form a
quartet for small values of the gain-loss parameter, while the third is
imaginary (i.e., the configuration is unstable due to the real parts of the
eigenvalues within the quartet). At $\gamma =1.17$, the eigenvalues of the
complex quartet collapse into two imaginary pairs, rendering the
configuration stable, in a narrow parametric interval. At $\gamma =1.24$,
the former imaginary pair becomes real, destabilizing the state again, while
subsequent bifurcations of imaginary pairs into real ones occur at $\gamma
=1.28$ and $\gamma =1.74$ (at the latter point, all three non-zero pairs
are real). Shortly thereafter, two of these pairs collide at $\gamma =1.76$
and rearrange into a complex quartet, which exists along with the real pair
past that point.

\item case 2 --- black squares. In contrast to all other branches, this one
is \textit{always} unstable. One of the two nonzero eigenvalue pairs is
always real (while the other is always imaginary), as seen in Fig.~\ref%
{figzpzm2}. This branch also terminates at the
exceptional point $\gamma =\pm 2k$%
, as relation $\sin \left( \phi _{b}\right) =-\gamma /\left( 2k\right) $
cannot hold at $|\gamma |>|2k|$. This branch collides with the two previous
ones via a very degenerate bifurcation (that could be dubbed a ``double
saddle-center" bifurcation), which involves 3 branches instead of two as in
the case of the generic saddle-center bifurcation, and two distinct
eigenvalue pairs colliding at the origin of the spectral plane.

%\item case 2 --- black squares\\
%In contrast to all other branches, this branch not only
%violates the $\cP\bT-$symmetry but it is, in fact, \textit{%
%always} unstable.
%% --- a feature which is also reflected
%%in the broken $\cP\bT-$symmetry of this stationary mode.
%One of the two nonzero eigenvalue pairs is always
%real (while the other is always imaginary), as seen in
%Fig.~\ref{figzpzm2}. This branch also terminates at the
%exceptional point $\g=\pm 2k$,
%as the relation $\sin \phi _{b}=-\gamma /\left( 2k\right)
%$ cannot hold at $|\gamma| >|2k|$. This branch collides with the two
%previous ones via a degenerate bifurcation (that could be
%dubbed a ``double
%saddle-center" bifurcation). The latter involves 3 branches instead of
%two, in the case of the generic saddle-center bifurcation, and two
%distinct eigenvalue pairs colliding at the origin of the spectral
%plane.
\end{itemize}

\begin{figure}[tph]
\subfigure[\ blue circle
branch]{\scalebox{0.4}{\includegraphics{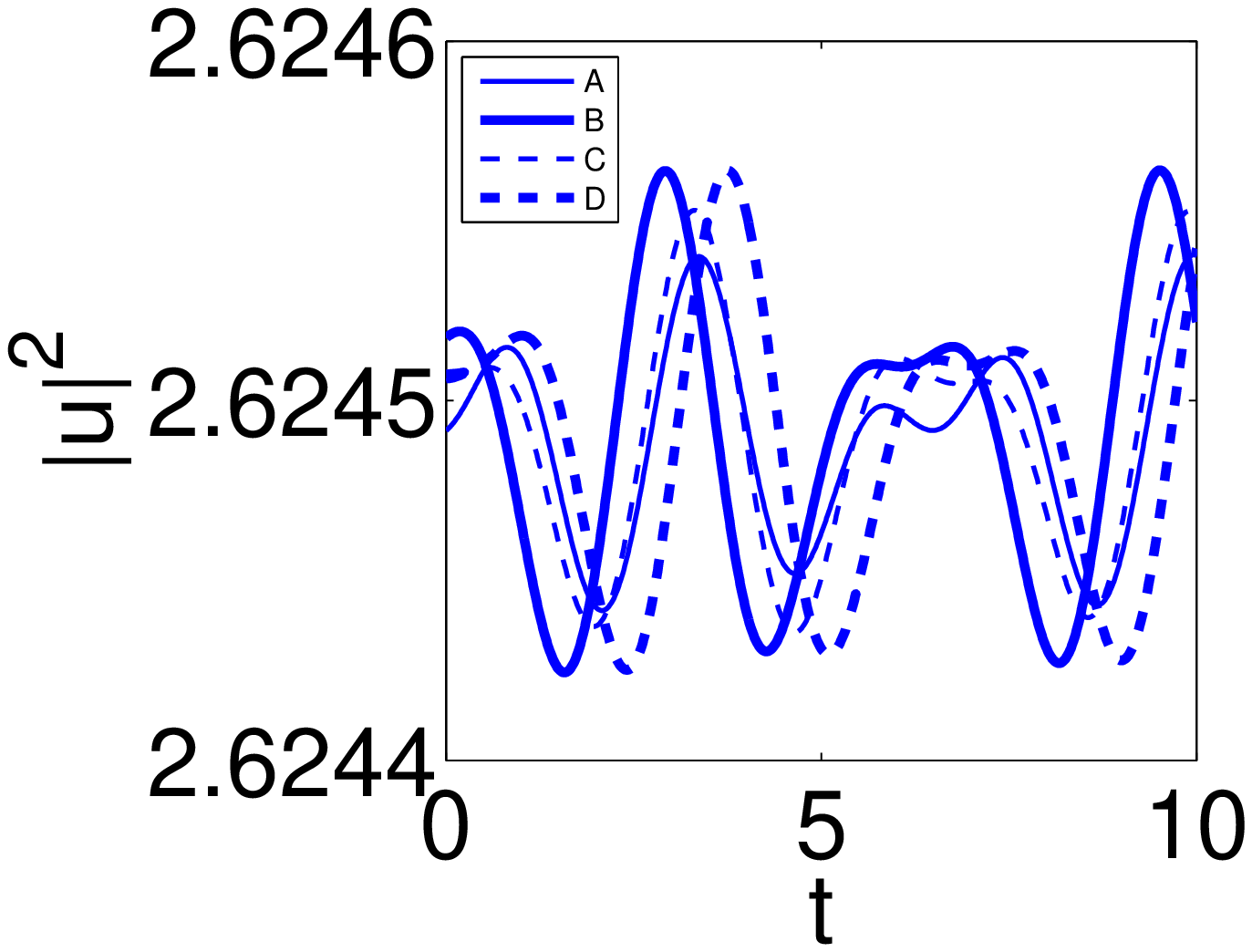}}}\hfill
\subfigure[\ red
cross branch]{\scalebox{0.4}{\includegraphics{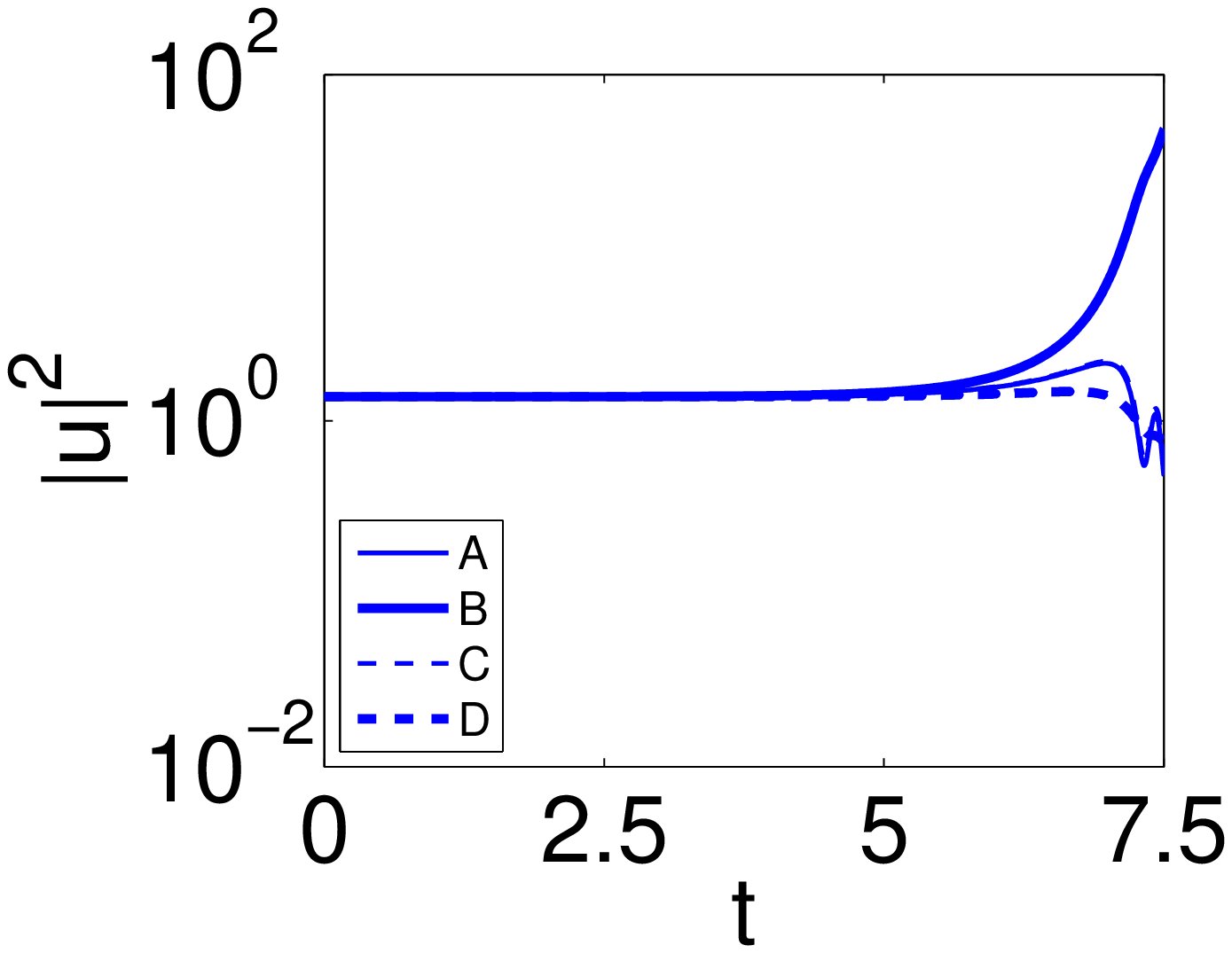}}}\newline
\subfigure[\
black square branch]{\scalebox{0.4}{\includegraphics{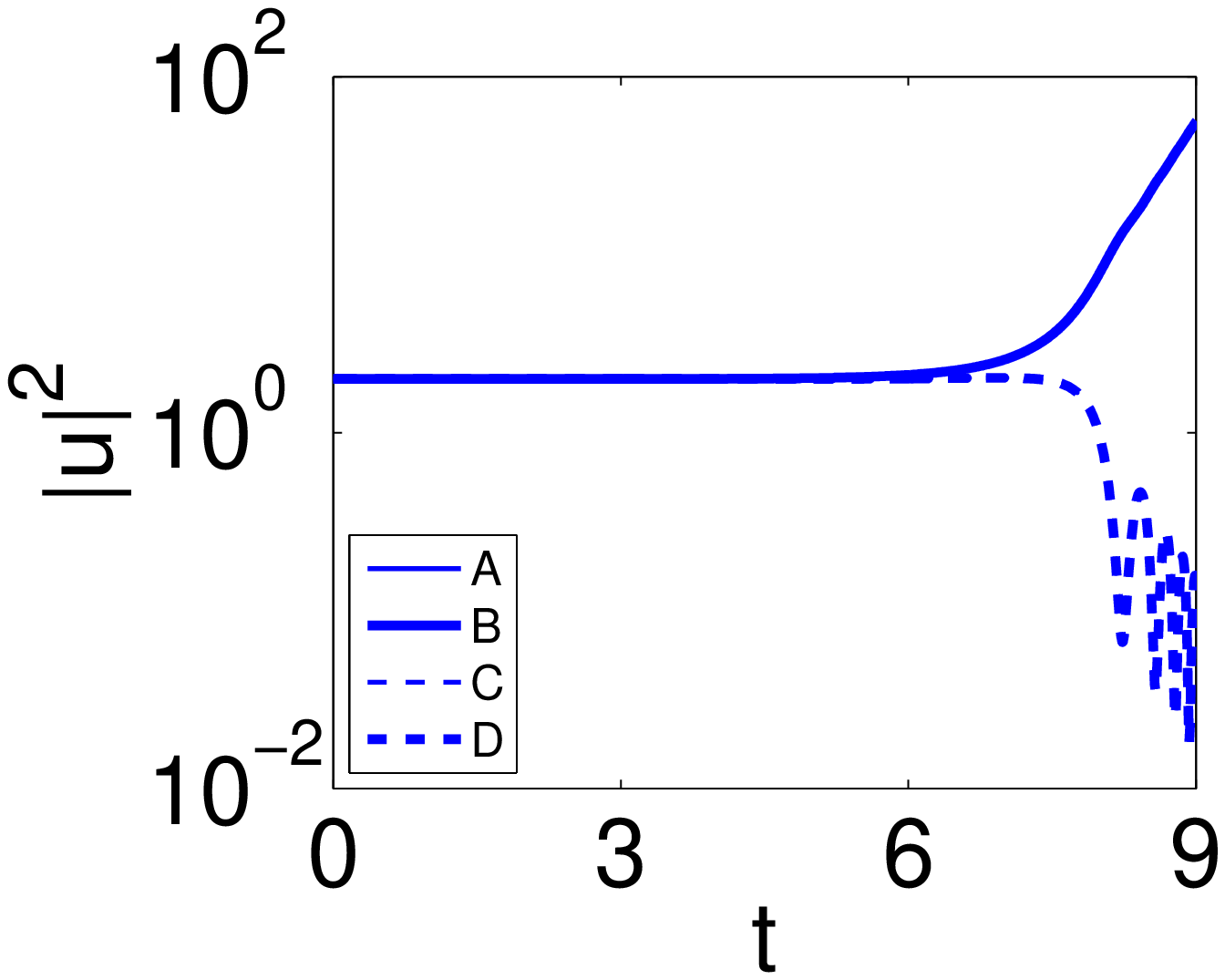}}}\hfill
\subfigure[\ green star
branch]{\scalebox{0.4}{\includegraphics{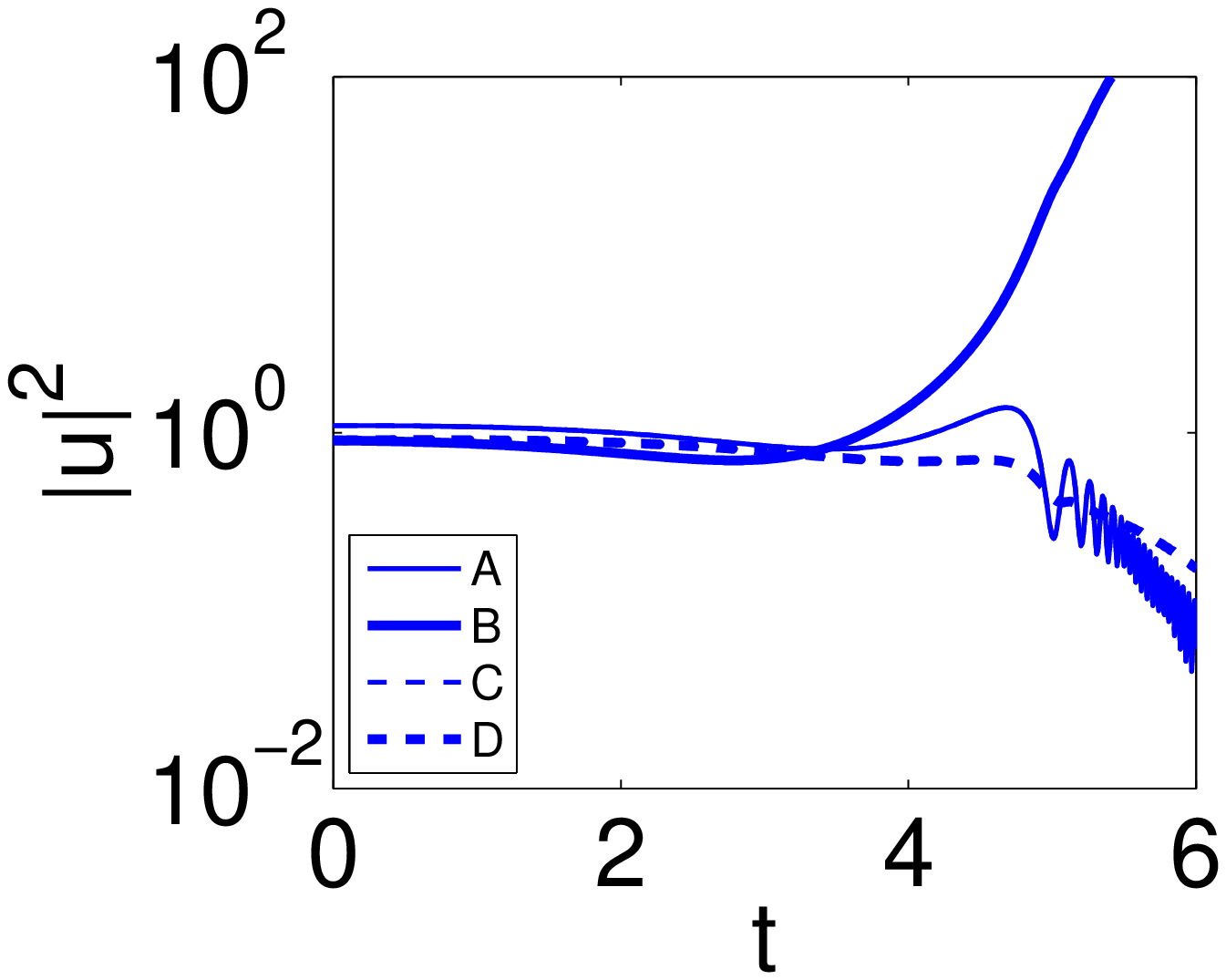}}}
\caption{(Color online) The perturbed evolution of different branches from
Figs. \protect\ref{figzpzm1} and \protect\ref{figzpzm2} at $\protect\gamma %
=1.9$. Thin solid, thick solid, thin dashed, and thick dashed curves
correspond to nodes A, B, C, D in Fig.~\protect\ref{modes}(a), respectively.
In panel (b), the plots pertaining to sites A and C [see Fig. \protect\ref%
{modes}(a)] overlap. Similarly, pairs of the plots for (A,B) and (C,D)
overlap in (c), and for (A,C) they overlap in (d).}
\label{stabzpzm}
\end{figure}

By means of direct simulations, we have also examined the dynamics of the
modes belonging to different branches in Fig.~\ref{stabzpzm}. The stable
blue-circle branch demonstrates only oscillations under perturbations. This
implies that, despite the presence of the gain-loss profile, none of the
perturbation eigenmodes grows in this case. Nevertheless, the three other
branches ultimately manifest their dynamical instability, which is observed
through the growth of the amplitude at the gain-carrying site [B, in Fig. %
\ref{modes}(a)] at the expense of the lossy site (D). That is, the amplitude
of the solution at the site with the gain grows, while the amplitude of the
solution at the dissipation site loses all of its initial power. Depending
on the particular solution, passive sites (the ones without gain or loss,
such as A and C) may be effectively driven by the gain (as in the case of
the black-square-branch, where the site A is eventually amplified due to the
growth of the amplitude at site B) or by the loss (red-cross and green-star
branches, where, eventually, the amplitudes at both A and C sites lose all
of their optical power).

\subsection{The plaquette of the +-+- type}

We now turn to the generalized (not exactly $\cP\bT$-symmetric)
configuration\footnote{For the terminology concerning  exact $\cP\bT-$symmetry, spontaneously broken $\cP\bT-$symmetry and completely broken $\cP\bT-$symmetry see the discussion of Eqs. \rf{a13} and \rf{a13-2}.} featuring the alternation of the gain and loss along the
plaquette in panel (b) of Fig.~\ref{modes}.
Indeed, the absence of  $\mathcal{PT}$-symmetry in this case is
mirrored in the existence of imaginary eigenvalues in the linear
problem of Eqs.~(\ref{a15-2}), as soon as $\gamma \neq 0$. The corresponding nonlinear solutions (with $E\not\in\RR$) are not covered by the stationary solution ansatz \rf{E}. Stationary solutions (with $E\in\RR$) solely belong to dynamical regimes below the concrete $\cP\bT-$thresholds.
Apart from the two $\cP\bT-$symmetry violating solutions, there should exist at least two stationary solutions which we construct in analogy to
[cf. Eqs. (\ref{zpzm2})] from
%\begin{eqnarray}
%i\dot{u}_{A} &=&-k(u_{B}+u_{D})-|u_{A}|^{2}u_{A}+i\gamma u_{A},  \notag \\
%i\dot{u}_{B} &=&-k(u_{A}+u_{C})-|u_{B}|^{2}u_{B}-i\gamma u_{B},  \notag \\
%i\dot{u}_{C} &=&-k(u_{B}+u_{D})-|u_{C}|^{2}u_{C}+i\gamma u_{C},  \notag \\
%i\dot{u}_{D} &=&-k(u_{A}+u_{C})-|u_{D}|^{2}u_{D}-i\gamma u_{D};
%\label{pmpm1}
%\end{eqnarray}%
\begin{eqnarray}
Ea &=&k(b+d)+|a|^{2}a-i\gamma a,  \notag \\
Eb &=&k(a+c)+|b|^{2}b+i\gamma b,  \notag \\
Ec &=&k(b+d)+|c|^{2}c-i\gamma c,  \notag \\
Ed &=&k(a+c)+|d|^{2}d+i\gamma d.  \label{pmpm2}
\end{eqnarray}

\begin{figure}[tph]
\scalebox{0.4}{\includegraphics{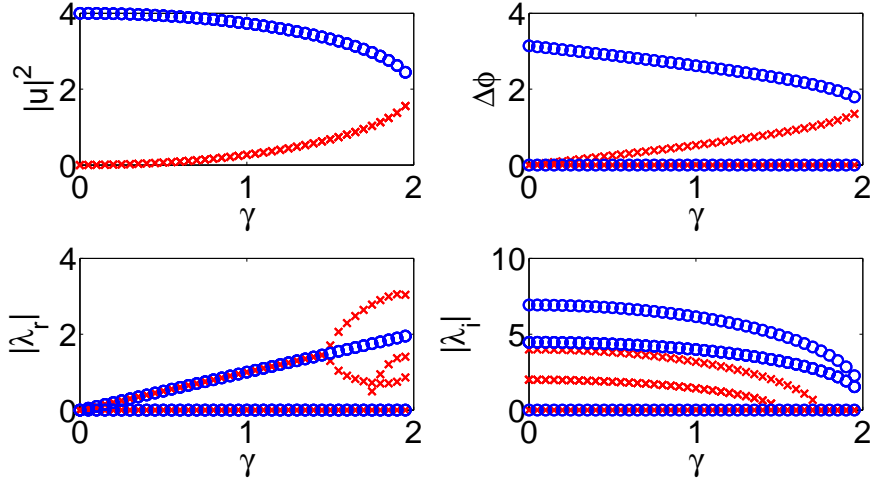}}
\caption{(Color online) The continuation of mode (\protect\ref{phiA}) and
its stability, supported by plaquette (b) \ in Fig. \protect\ref{modes}, for
$E=2$ and $k=1$.}
\label{figpmpm2}
\end{figure}

\begin{figure}[htp]
\label{figpmpm3}\scalebox{0.4}{\includegraphics{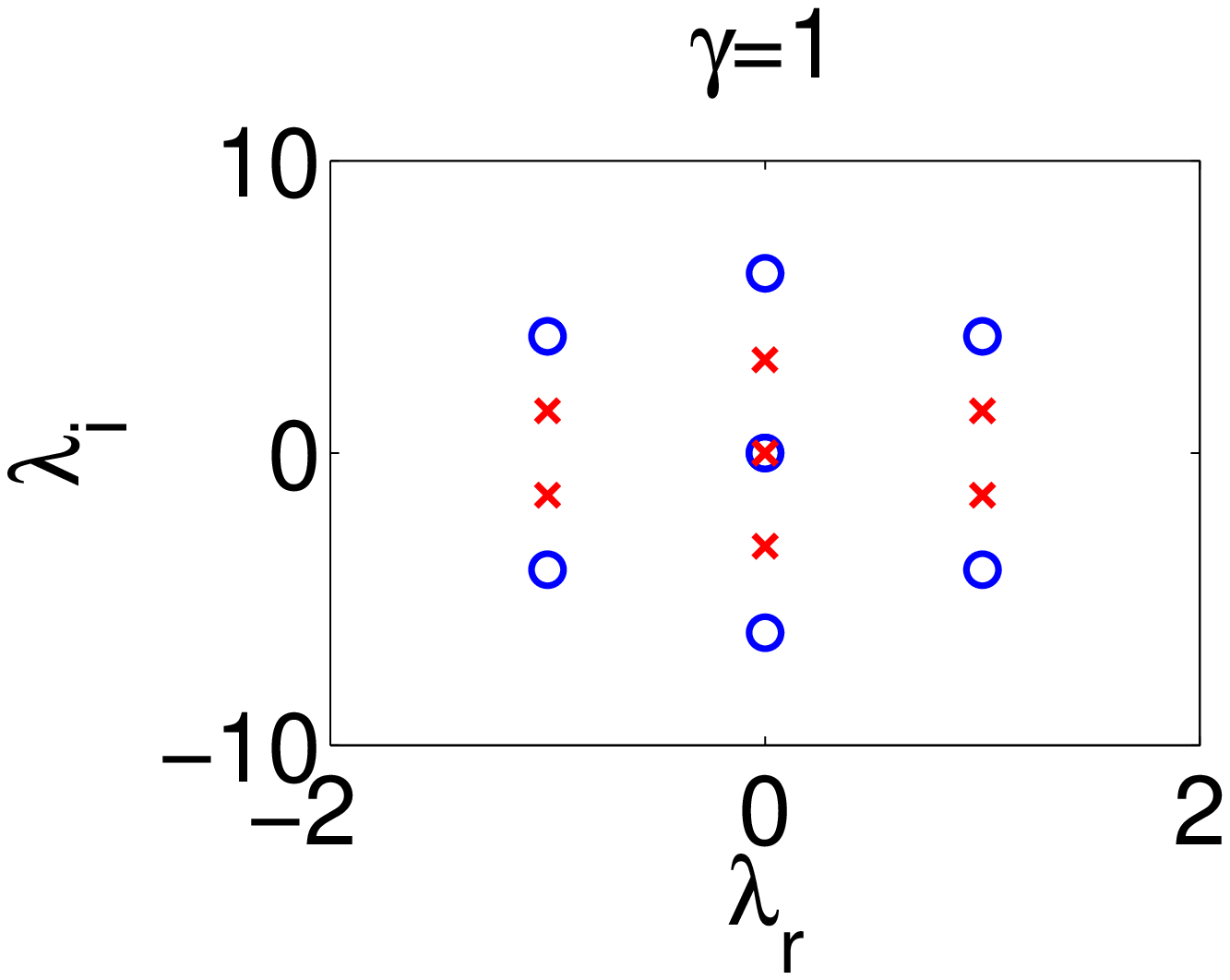}} %
\scalebox{0.4}{\includegraphics{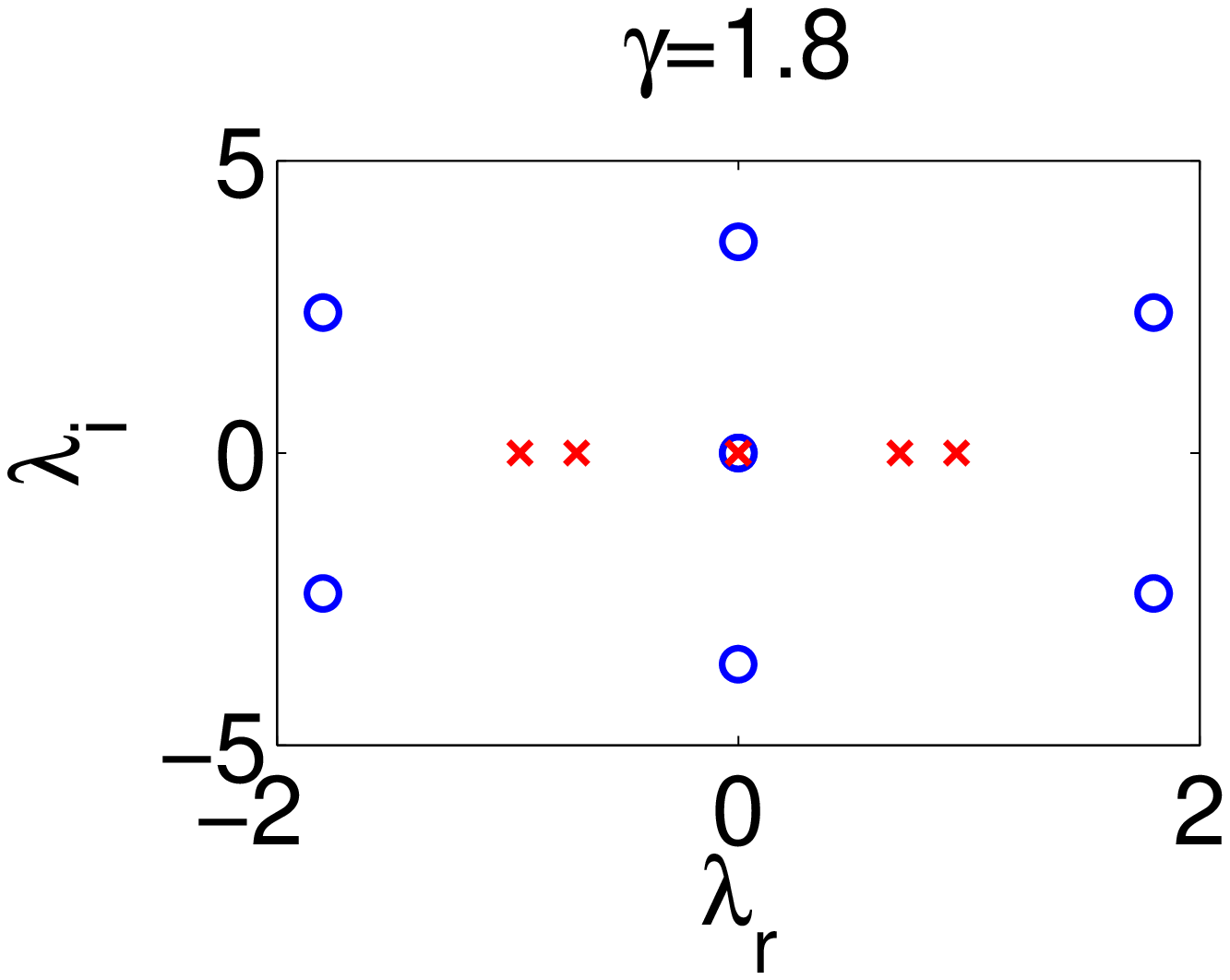}}
\caption{Two typical stability plots for branch (\protect\ref{phiA}), for $%
E=2$, $k=1$ and $\protect\gamma =1$ and $1.8$, respectively.}
%When $\gamma$ is small, there are one pair of purely imaginary eigenvalues
%and four complex eigenvalues forming quartet. Those four collide into two pairs of
%real eigenvalues at $\gamma=1.5$. After that the purely imaginary eigenvalues collide
%into a pair of real eigenvalues at $\gamma=1.74$.}
\end{figure}

\begin{figure}[tph]
\subfigure[\ blue circles
branch]{\scalebox{0.4}{\includegraphics{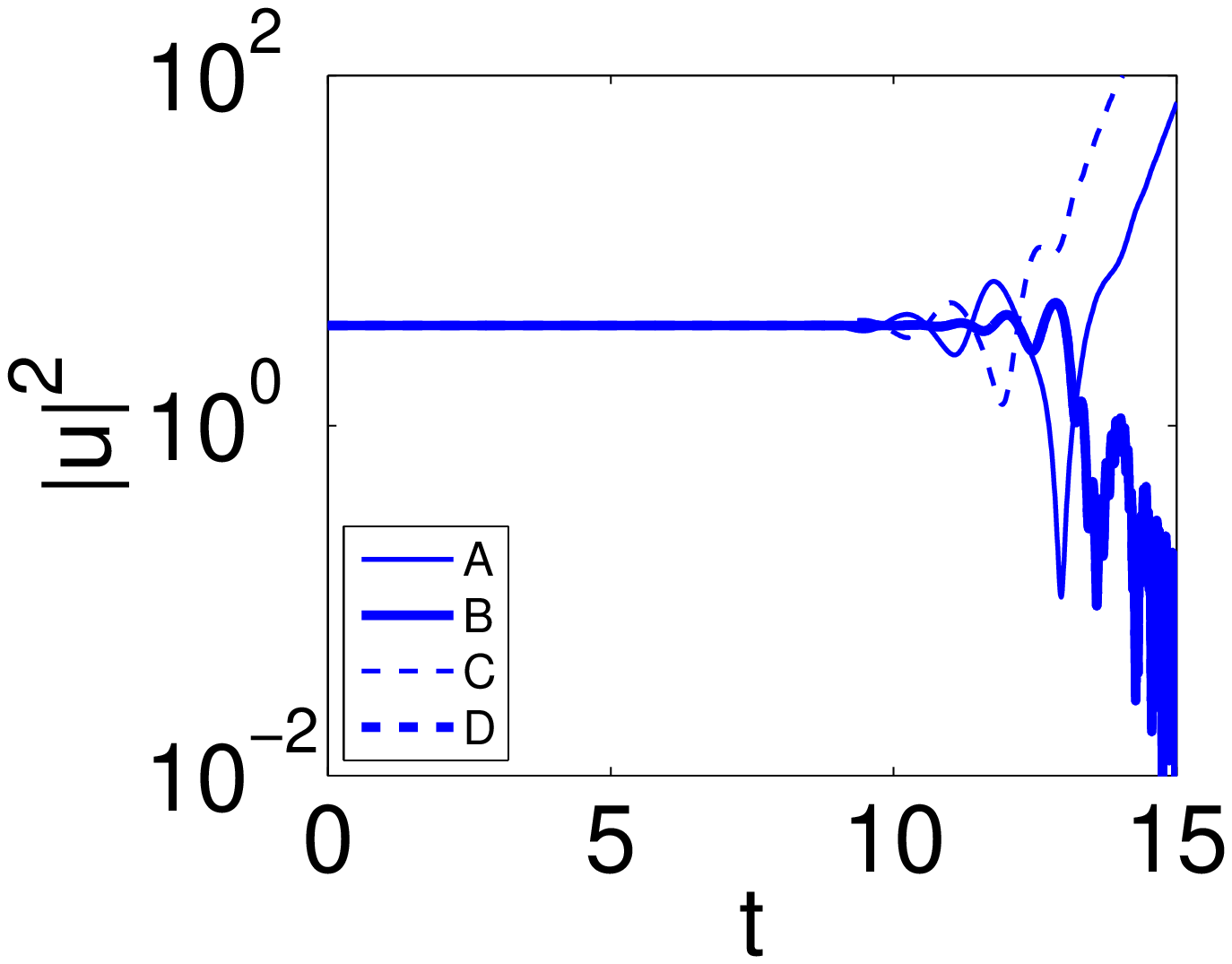}}}
\subfigure[\ red
crosses branch]{\scalebox{0.4}{\includegraphics{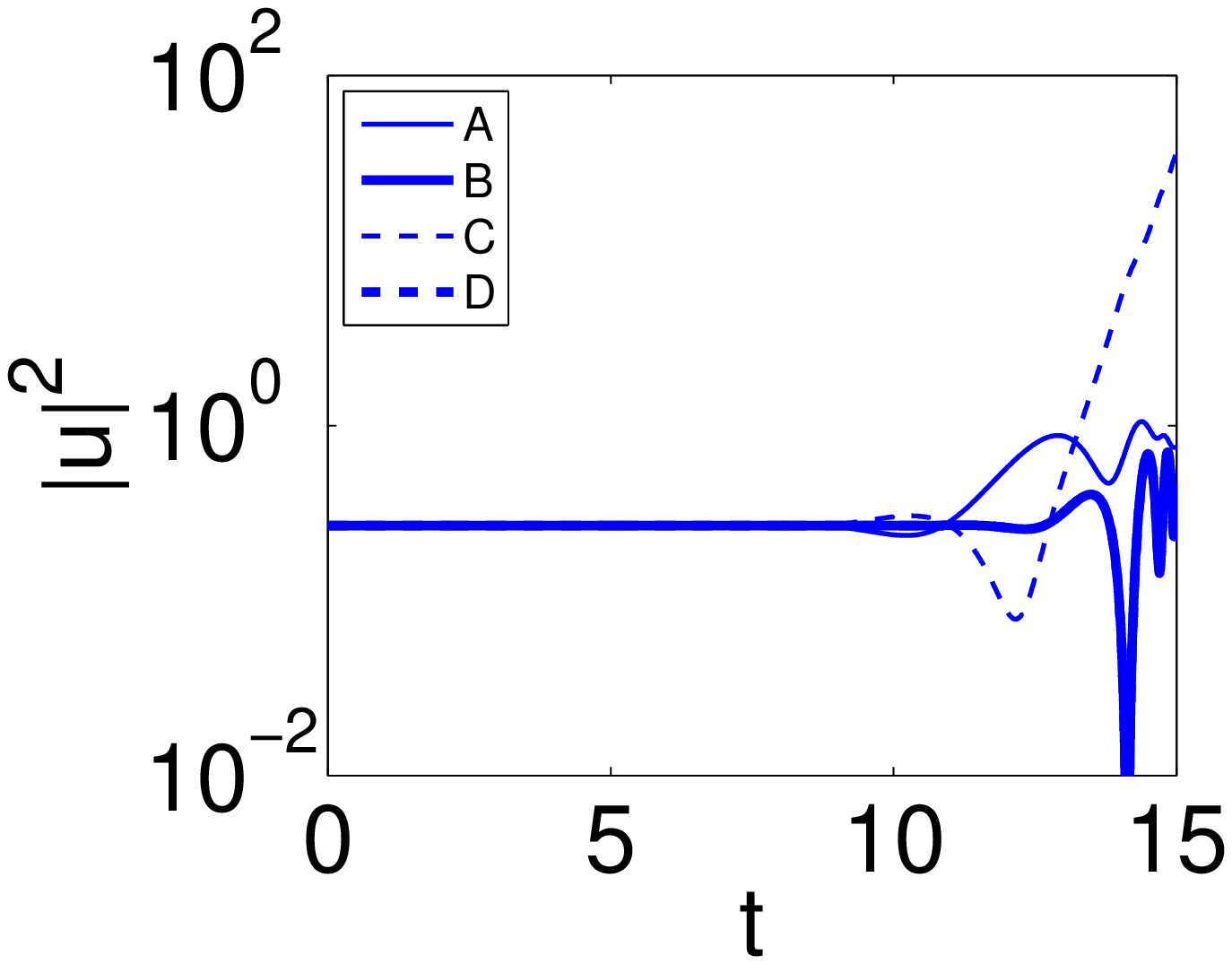}}}
\caption{The perturbed evolution of the modes of type (\protect\ref{phiA})
at $\protect\gamma =1$ corresponding to the left panel of Fig. \protect\ref%
{figpmpm3}. The plots pertaining to sites B and D [see Fig. \protect\ref%
{modes}(b)] overlap in both panels.}
\label{stabpmpm}
\end{figure}

%Plug the polar representation of the form $a=Ae^{i\phi_a}, b=Be^{i\phi_b},
%c=Ce^{i\phi_c}, d=De^{i\phi_d}$ into the above equations, and for simplicity, we
Substituting the Madelung representation (\ref{aA}) and setting $A=B=C=D$
(for illustration purposes, we focus here only on this simplest case),
we obtain
\begin{eqnarray}
\sin (\phi _{b}-\phi _{a})+\sin (\phi _{d}-\phi _{a}) &=&\sin (\phi
_{b}-\phi _{c})+\sin (\phi _{d}-\phi _{c})=\frac{\gamma }{k},
\label{phi1}
\\
\cos (\phi _{b}-\phi _{a})+\cos (\phi _{d}-\phi _{a}) &=&\cos (\phi
_{b}-\phi _{c})+\cos (\phi _{d}-\phi _{c})=\frac{E-A^{2}}{k}.
\label{phi2}
\end{eqnarray}%
Further, fixing $\phi _{a}=\phi _{c}=0$, Eqs. (\ref{phi1}) and (\ref{phi2})
yield
\begin{equation}
\sin \phi _{b}=\sin \phi _{d}=\frac{\gamma }{2k},\ A^{2}=E\pm \sqrt{%
4k^{2}-\gamma ^{2}}.  \label{phiA}
\end{equation}%
%
%
%
%
%We consider the following two distinct possibilities:
%\begin{enumerate}
%\item $A^2=E,\quad k=\gamma/2$: Both of the phase and amplitude are fixed in this case. Set
%$\phi_a=\phi_c=0$, we have $\phi_b=\phi_d=\frac{\pi}{2}$. See Fig.~\ref{figpmpm1}.
%\begin{figure}[htp]
%\scalebox{0.5}{\includegraphics{pmpm1.eps}}
%\scalebox{0.4}{\includegraphics{pmpm_stability11.eps}}
%\scalebox{0.4}{\includegraphics{pmpm_u1r1.eps}}
%\caption{The solution profile in case 1 of mode +-+- with $E=1$ and $k=1$.
%In the time evolution panel (bottom right), B and D overlap.}
%\label{figpmpm1}
%\end{figure}
Obviously, the solution terminates at point $\gamma =\pm 2k$. Similar to
what was done above, the continuation of this branch and typical examples of
its linear stability are shown in Figs.~\ref{figpmpm2} and \ref{figpmpm3},
respectively. From here it is seen that the blue-circle branch, which has a
complex quartet of eigenvalues, is always unstable. In fact, the gain-loss
alternating configuration is generally found to be more prone to the
instability. The red-cross branch is also unstable via a similar complex
quartet of eigenvalues. This quartet, however, breaks into two real pairs
for $\gamma \geq 1.5$, and, eventually, the
additional imaginary eigenvalue pair becomes real too at $\gamma >1.74$,
making the solution highly unstable with three real eigenvalue pairs. The
manifestation of the instability is shown in Fig.~\ref{stabpmpm}, typically
amounting to the growth of the amplitudes at one or more gain-carrying
sites. %The instability
%may [as in case (b)] or may not [case (c)] induce the growth also at lossy
%sites.

%\end{enumerate}

\subsection{The plaquette of the ++- - type}

\begin{figure}[htp]
\label{figppmm}\scalebox{0.4}{\includegraphics{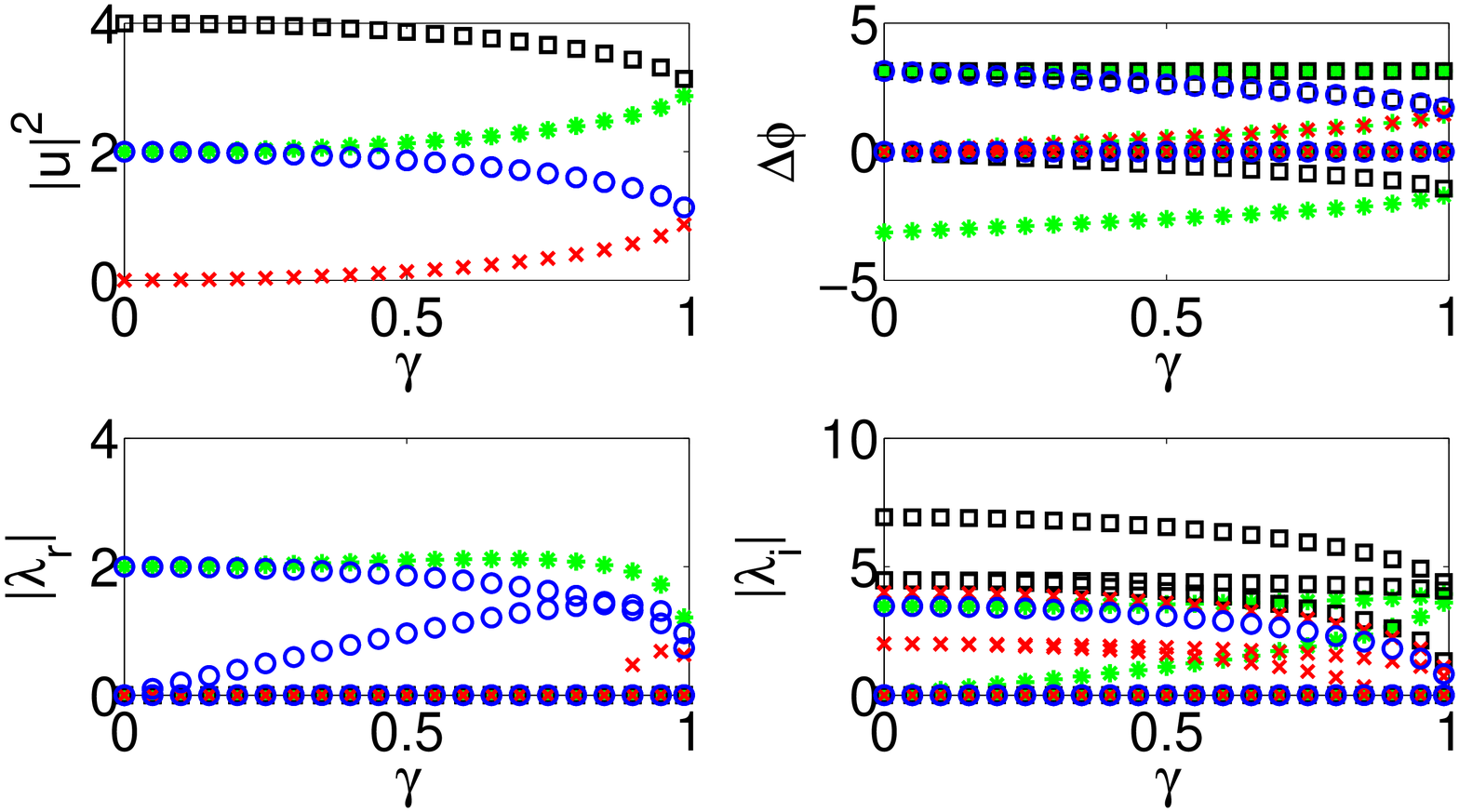}} %
\scalebox{0.4}{\includegraphics{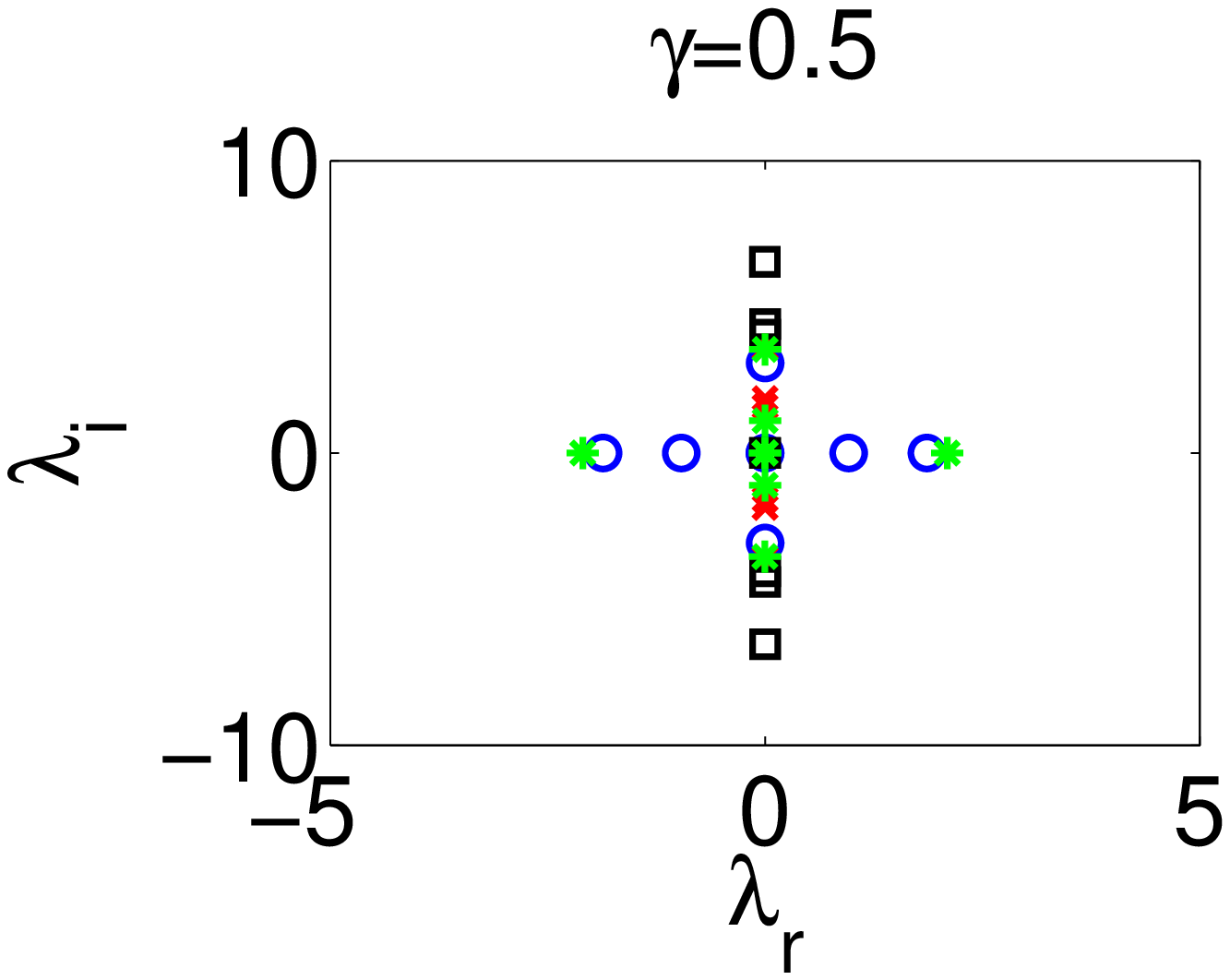}} \scalebox{0.4}{%
\includegraphics{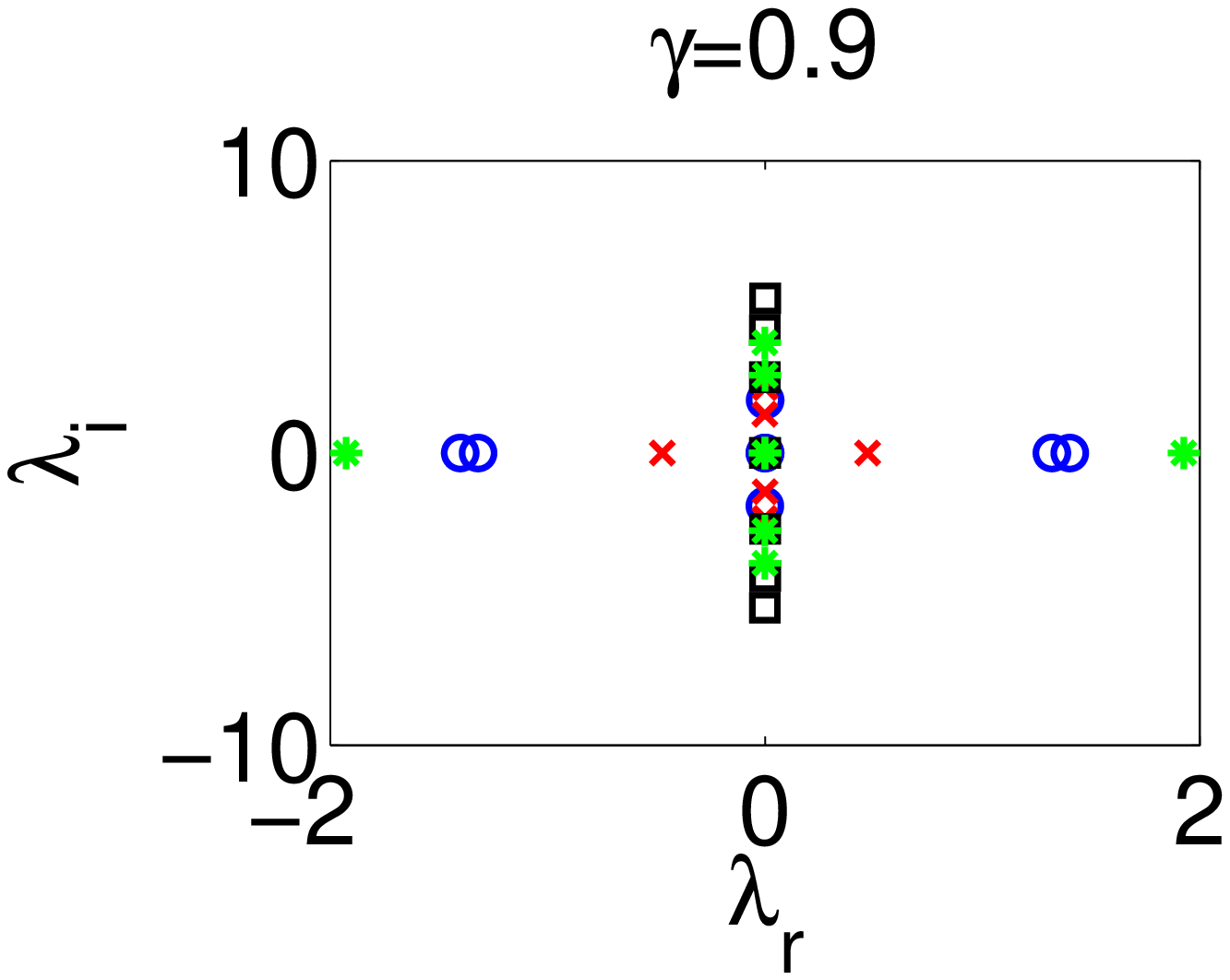}}
\caption{(Color online) The characteristics of the mode of the ++- - type,
supported by plaquette (c) in Fig. \protect\ref{modes}, and given in
analytical form by Eqs. (\protect\ref{+-1})- (\protect\ref{+-4}), for $E=2$
and $k=1$. The blue circles correspond to the completely unstable branch
with the upper sign in Eq. (\protect\ref{+-1}), while the red crosses
pertain to branch with the lower sign, which is stable at $\protect\gamma %
<0.86$. The black-square and green-star branches correspond to the upper and
lower sign in Eq. (\protect\ref{+-3}), respectively. The former one is
always stable, while the later one is always unstable. All four branches
terminate at the critical point $|\protect\gamma |=|k|$ of the linear $%
\mathcal{PT}$-symmetric system.}
%The circled blue one stands for the branch where $A^2=E-k+\sqrt{k^2-\gamma^2}$, and
%the crossed red one stands for $A^2=E-k-\sqrt{k^2-\gamma^2}$.
%The circled blue branch has two pairs of real and one pair of purely imaginary
%eigenvalues. The crossed red one has three pairs of purely imaginary eigenvalues,
%one of which collide into a pair of real eigenvalues at $\gamma=0.86$.
%The two branches both terminate at $\gamma=k$.}
\end{figure}

\begin{figure}[tph]
\subfigure[\ blue circles
branch]{\scalebox{0.4}{\includegraphics{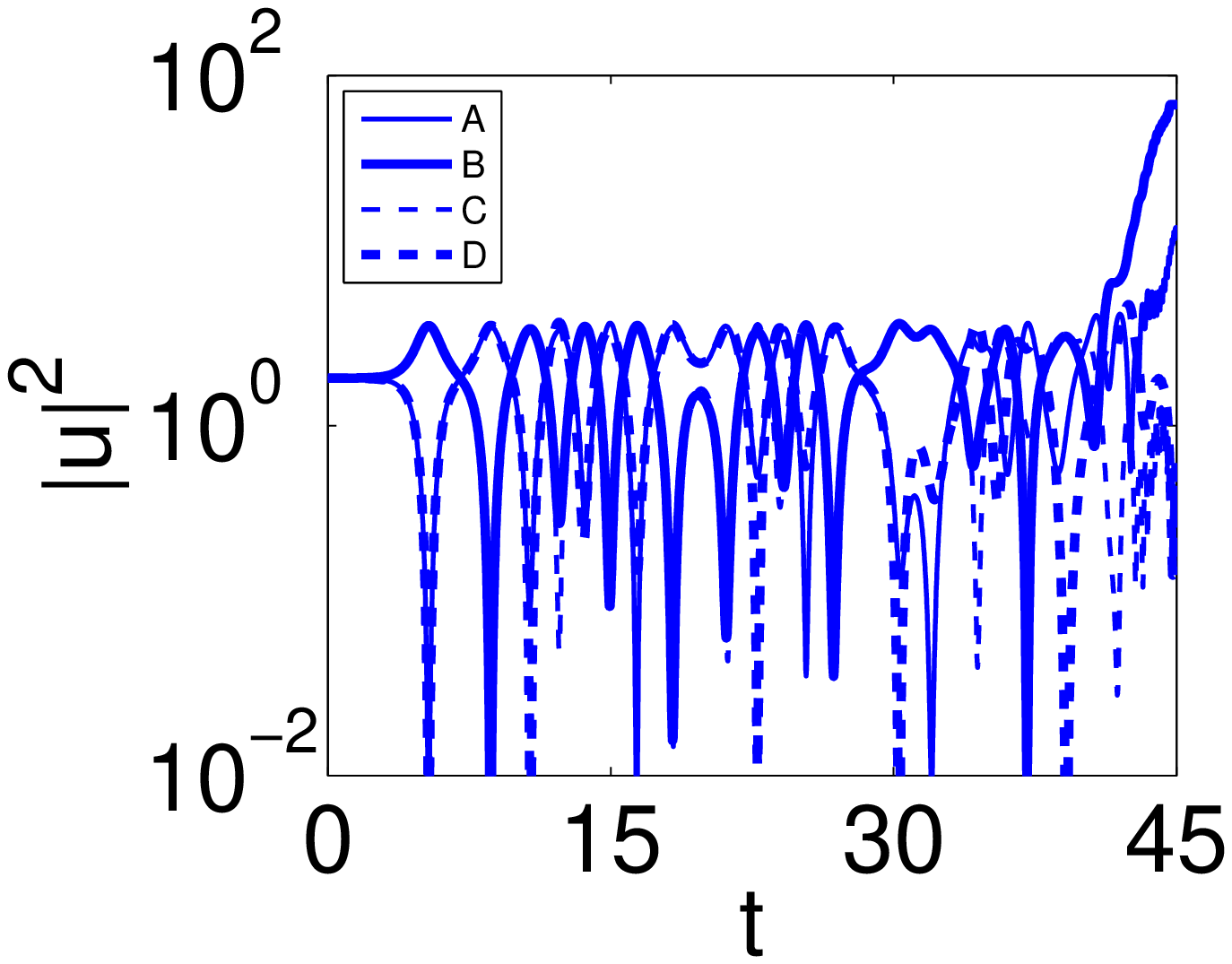}}}\hfill
\subfigure[\
red crosses branch]{\scalebox{0.4}{\includegraphics{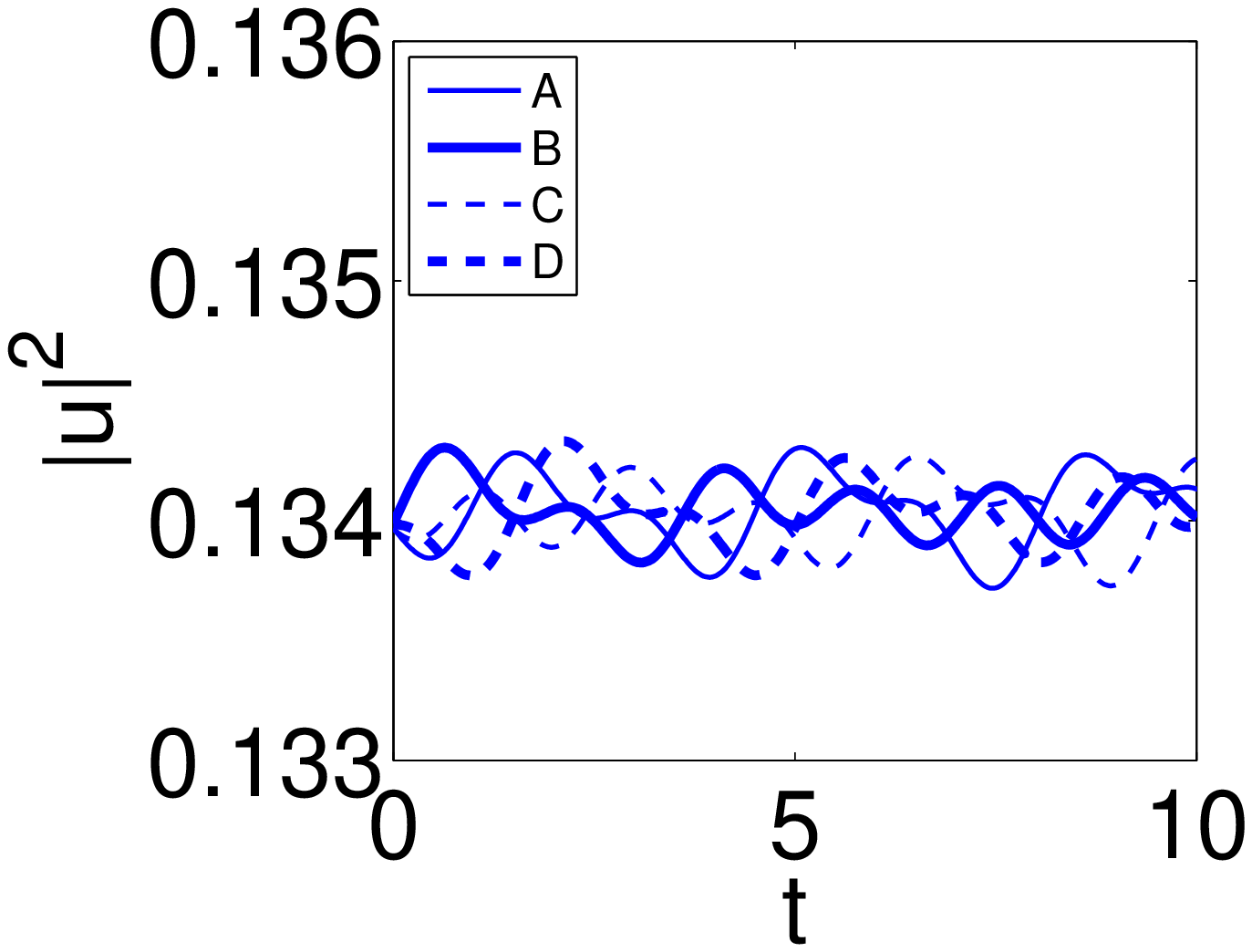}}} \newline
\subfigure[\ black squares
branch]{\scalebox{0.4}{\includegraphics{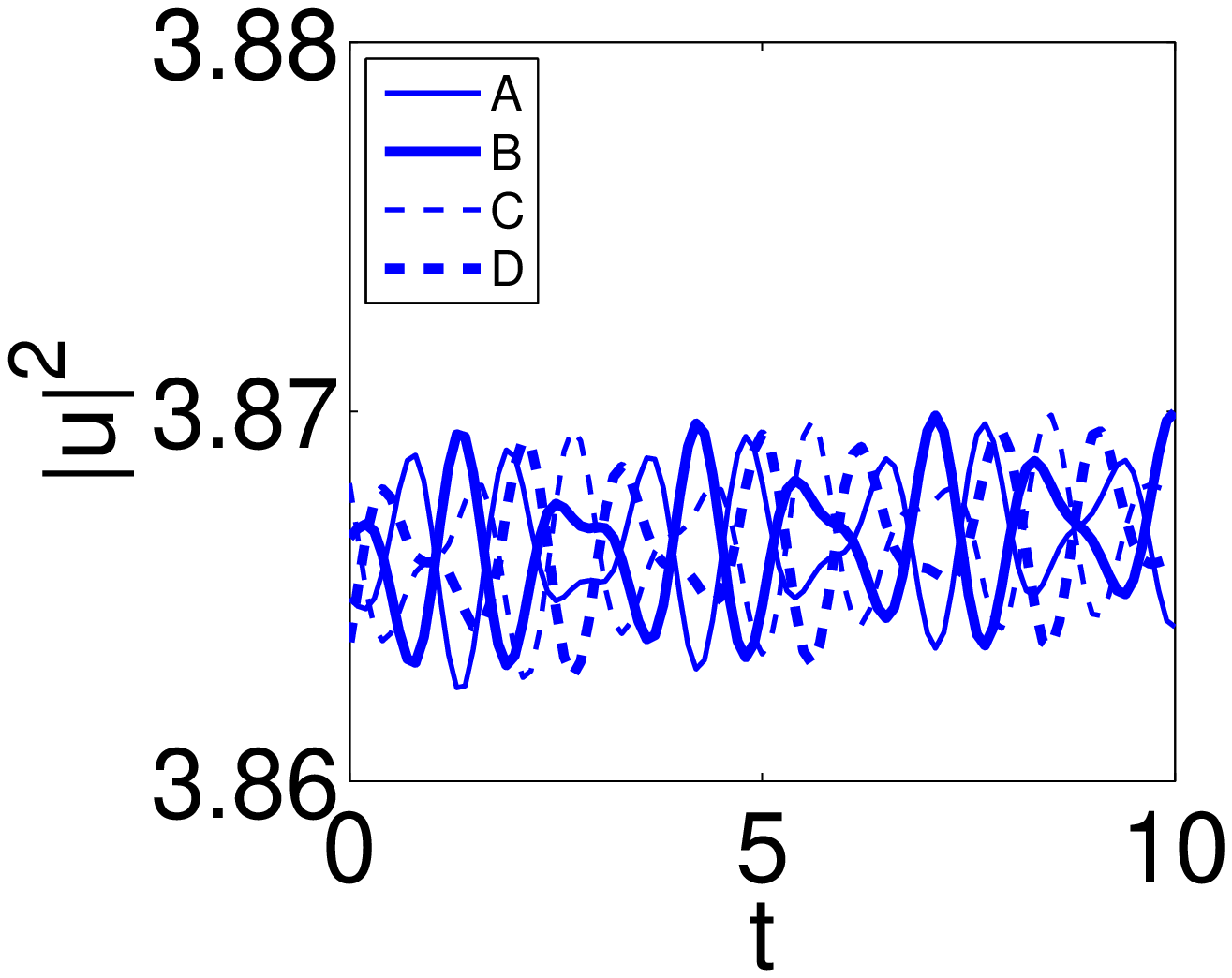}}}\hfill
\subfigure[\
green stars branch]{\scalebox{0.4}{\includegraphics{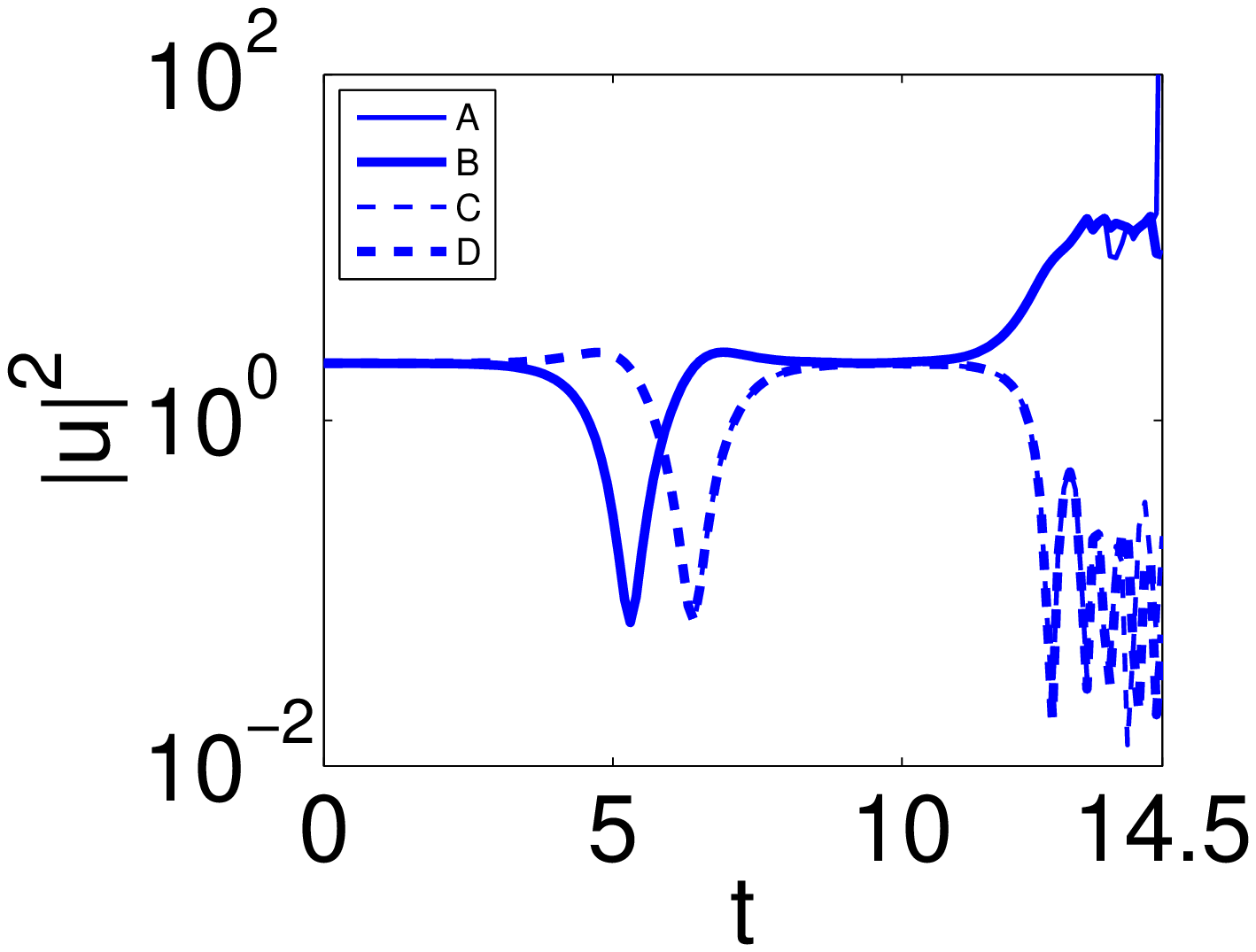}}}
\caption{(Color online) The perturbed evolution of the four branches of the
analytical solutions given by Eqs. (\protect\ref{+-1})-(\protect\ref{+-4}),
which correspond to Fig.~\protect\ref{figppmm} with $\protect\gamma =0.5$.}
\label{stabppmm}
\end{figure}

We now turn to the plaquette in Fig. \ref{modes}(c), which involves parallel
rows of gain and loss. In this case,
%the dynamical and
the stationary equations
are
%\begin{eqnarray}
%i\dot{u}_{A} &=&-k(u_{B}+u_{D})-|u_{A}|^{2}u_{A}+i\gamma u_{A},  \notag \\
%i\dot{u}_{B} &=&-k(u_{A}+u_{C})-|u_{B}|^{2}u_{B}+i\gamma u_{B},  \notag \\
%i\dot{u}_{C} &=&-k(u_{B}+u_{D})-|u_{C}|^{2}u_{C}-i\gamma u_{C},  \notag \\
%i\dot{u}_{D} &=&-k(u_{A}+u_{C})-|u_{D}|^{2}u_{D}-i\gamma u_{D};
%\label{ppmm1}
%\end{eqnarray}%
\begin{eqnarray}
Ea &=&k(b+d)+|a|^{2}a-i\gamma a,  \notag \\
Eb &=&k(a+c)+|b|^{2}b-i\gamma b,  \notag \\
Ec &=&k(b+d)+|c|^{2}c+i\gamma c,  \notag \\
Ed &=&k(a+c)+|d|^{2}d+i\gamma d.  \label{ppmm2}
\end{eqnarray}%
In this case too, we focus on symmetric states of the form of $A=B=C=D$ [see
Eq. (\ref{aA})], which gives rise to two solutions displayed in Fig.~\ref%
{figppmm}, represented by the following analytical solutions:%
%Similar as the mode +-+-, we now have two solutions:
%\begin{eqnarray}
%\sin(\phi_b-\phi_a)+\sin(\phi_d-\phi_a)&=&-\sin(\phi_b-\phi_c)-\sin(\phi_d-\phi_c)=\frac{\gamma}{k} \\
%\cos(\phi_b-\phi_a)+\cos(\phi_d-\phi_a)&=&\cos(\phi_b-\phi_c)+\cos(\phi_d-\phi_c)=\frac{E-A^2}{k},
%\end{eqnarray}
%which gives us
%\begin{equation}
%\phi_a=\phi_b,\quad \phi_c=\phi_d.
%\end{equation}

\begin{eqnarray}  \label{phiAquasi1D}
A^{2} &=&E-k\pm \sqrt{k^{2}-\gamma ^{2}},  \label{+-1} \\
\phi _{a} &=&\phi _{b}=0,~\sin \phi _{c}=\sin \phi _{d}=\frac{\gamma }{k};
\label{+-2}
\end{eqnarray}
\begin{eqnarray}  \label{phiAquasi1D2}
A^{2} &=&E+k\pm \sqrt{k^{2}-\gamma ^{2}},  \label{+-3} \\
\phi _{a} &=&0,~\phi _{b}=\pi ,~\phi _{c}=\phi _{d}-\pi ,~\sin \phi _{d}=%
\frac{\gamma }{k},  \label{+-4}
\end{eqnarray}%
The analysis demonstrates that the branch with the upper sign in Eq. (\ref%
{+-1}) is always unstable (through two real pairs of eigenvalues), as shown
by blue circles in Fig.~\ref{figppmm}. On the other hand, the branch denoted
by the red crosses, which corresponds to the lower sign in Eq. (\ref{+-1})
is stable up to $\gamma =0.86$, and then it gets unstable through a real
eigenvalue pair. The black-squares branch with the upper sign in Eq.~(\ref%
{+-3}) is always stable, while the green-star branch with the lower sign in
Eq.~(\ref{+-3}) is always unstable. At the linear-$\mathcal{PT}$-symmetry
breaking point $\gamma =k$, we observe a strong degeneracy, since all the
three pairs of eigenvalues for two of the branches (in the case of the blue
circles, two real and one imaginary, and in the case of red crosses--- one
real and two imaginary) collapse at the origin of the spectral plane. On the
other hand, the black-squares branch is always stable with three imaginary
eigenvalue pairs, while the green-star branch has two imaginary and one real
pair of eigenvalues. Between the latter two, there is again a collision of a
pair at the origin at the critical condition, $\gamma =k$. Direct
simulations, presented for $\gamma =0.5$ in Fig.~{\ref{stabppmm},}
demonstrate the stability of the lower-sign black-squares branch, while the
instability of the waveform associated with the blue circles and the green
stars leads to the growth and decay of the amplitudes at the sites carrying,
respectively, the gain and loss. Notice that at the parameter values
considered here, the red-cross branch is also dynamically stable as shown in
the top right panel of Fig.~{\ref{stabppmm}}.

%The circled blue one stands for the branch where $A^2=E-k+\sqrt{k^2-\gamma^2}$, and
%the crossed red one stands for $A^2=E-k-\sqrt{k^2-\gamma^2}$.
%The circled blue branch has two pairs of real and one pair of purely imaginary
%eigenvalues. The crossed red one has three pairs of purely imaginary eigenvalues,
%one of which collide into a pair of real eigenvalues at $\gamma=0.86$.
%The two branches both terminate at $\gamma=k$.}

\subsection{The plaquette of the +-0+- type}

Lastly, motivated by the existence of known ``cross"-shaped discrete-vortex
modes in 2D conservative lattices, in addition to the fundamental discrete
solitons~\cite{pgk_book,malomed}, we have also examined the five-site
configuration, in which the central site does not carry any gain or loss,
while the other four feature a $\mathcal{PT}$-balanced distribution of the
gain and loss, as shown in panel (d) of Fig.~\ref{modes}.
%In this case, the
%dynamical equations are
%\begin{eqnarray}
%i\dot{u}_{A} &=&-ku_{C}-|u_{A}|^{2}u_{A}-i\gamma u_{A},  \notag \\
%i\dot{u}_{B} &=&-ku_{C}-|u_{B}|^{2}u_{B}+i\gamma u_{B},  \notag \\
%i\dot{u}_{C} &=&-k(u_{A}+u_{B}+u_{D}+u_{E})-|u_{C}|^{2}u_{C},  \notag \\
%i\dot{u}_{D} &=&-ku_{C}-|u_{D}|^{2}u_{D}-i\gamma u_{D},  \notag \\
%i\dot{u}_{E} &=&-ku_{C}-|u_{E}|^{2}u_{E}+i\gamma u_{E}.  \label{pmzpm1}
%\end{eqnarray}%
Seeking for stationary states with propagation constant, $G$
[instead of $E$ in Eq. (\ref{E}), as in this case we reserve label
$E$ for one of the sites of the 5-site plaquette in Fig.
\ref{modes}(d)], we get:
\begin{eqnarray}
Ga &=&kc+|a|^{2}a+i\gamma a,  \notag \\
Gb &=&kc+|b|^{2}b-i\gamma b,  \notag \\
Gc &=&k(a+b+d+e)+|c|^{2}c,  \notag \\
Gd &=&kc+|d|^{2}d+i\gamma d,  \notag \\
Ge &=&kc+|e|^{2}e-i\gamma e.  \label{pmzpm2}
\end{eqnarray}%
Similarly as before, we use the Madelung decomposition $a=Ae^{i\phi
_{a}},b=Be^{i\phi _{b}},c=Ce^{i\phi _{c}},d=De^{i\phi _{d}},e=Ed^{i\phi
_{e}} $, cf. Eq. (\ref{aA}), and focus on the simplest symmetric solutions
with $A=B=D=E$. Without the loss of generality, we set $\phi _{c}=0$,
reducing the equations to

\begin{eqnarray}  \label{G}
C^{2}(G-C^{2})=4A^{2}(G-A^{2}),  \notag \\
(kC)^{2}=(\gamma A)^{2}+(GA-A^{3})^{2},  \notag \\
\sin \phi _{a}=\frac{\gamma A}{kC},  \notag \\
\phi _{a}=-\phi _{b}=\phi _{d}=-\phi _{e}.
\end{eqnarray}

\begin{figure}[htp]
\label{figpmzpm1}\scalebox{0.4}{\includegraphics{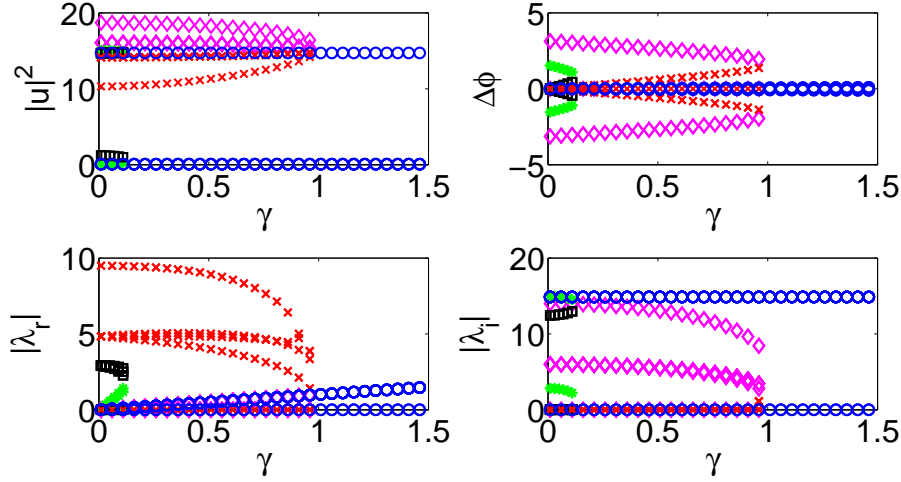}}
\caption{(Color online) The characteristics of the different branches of
solutions in the case of the five-site plaquette (d) in Fig. \protect\ref%
{modes} are shown for $G=15$ and $k=1$. The branches represented by the
chains of black squares and green stars terminate at $\protect\gamma =0.13$.
The branches depicted by red crosses and magenta diamonds terminate at $%
\protect\gamma =1$ [i.e., at the exceptional point $|\protect\gamma |=|k|$],
while the branch formed by blue circles continues past that point.}
%In the top left panel, there are two little curves of squared black and stared green that
%show $A$ around $15$ and $C$ around $0.6$. They terminate at $\gamma=0.13$.
%In the circled blue branch, $A$ stays around $0.065$ (not zero) and $C$ stays around
%$14.735$. Both of the crossed red and diamonded magenta branches terminates at $\gamma=1$,
%while the circled blue branch always survives.
%In the bottom left panel, the diamonded magenta shares the same line with the circled blue.}
\end{figure}

We report here numerical results for parameters $G=15,\ k=1$ (smaller $G$
yields similar results but with fewer solution branches). We have identified
five different solutions in this case, see Figs.~\ref{figpmzpm1} and \ref%
{figpmzpm2} for the representation of the continuation of the different
branches, and for typical examples of their stability (the latter is shown
for $\gamma =0.1$, $0.5$ and $0.95$). There are two branches (green stars
and black squares) that only exist at $\gamma <0.13,$ colliding and
terminating at that point. One of them has three real eigenvalue pairs and
one imaginary pair, while the other branch has two real and two imaginary
pairs. Two real pairs and one imaginary pair of green stars collide with two
real pairs and one imaginary pair of black squares, respectively, while the
final pairs of the two branches (one imaginary for the green stars and one
real for the black squares) collide at the origin of the spectral plane.
These collisions take place at $\gamma =0.13$, accounting for the
saddle-center bifurcation at the point where those two branches terminate.
On the other hand, there exist two more branches (red crosses and magenta
diamonds in Fig.~\ref{figpmzpm1}), which collide at $|\gamma |=|k|$. One of
these branches (the less unstable one, represented by magenta diamonds)
bears only an instability induced by an eigenvalue quartet, while the highly
unstable branch depicted by the red crosses has four real pairs (two of
which collide on the real axis and become complex at $\gamma >0.92$). Last
but not least, the blue circles
branch does not terminate at $\gamma =\pm k$, but
continues to larger values of the gain-loss parameter, $|\gamma |>|k|$. It
is also unstable (as the one represented by the magenta diamonds) due to a
complex quartet of eigenvalues.

\begin{figure}[tph]
\subfigure[\ blue circles
branch]{\scalebox{0.4}{\includegraphics{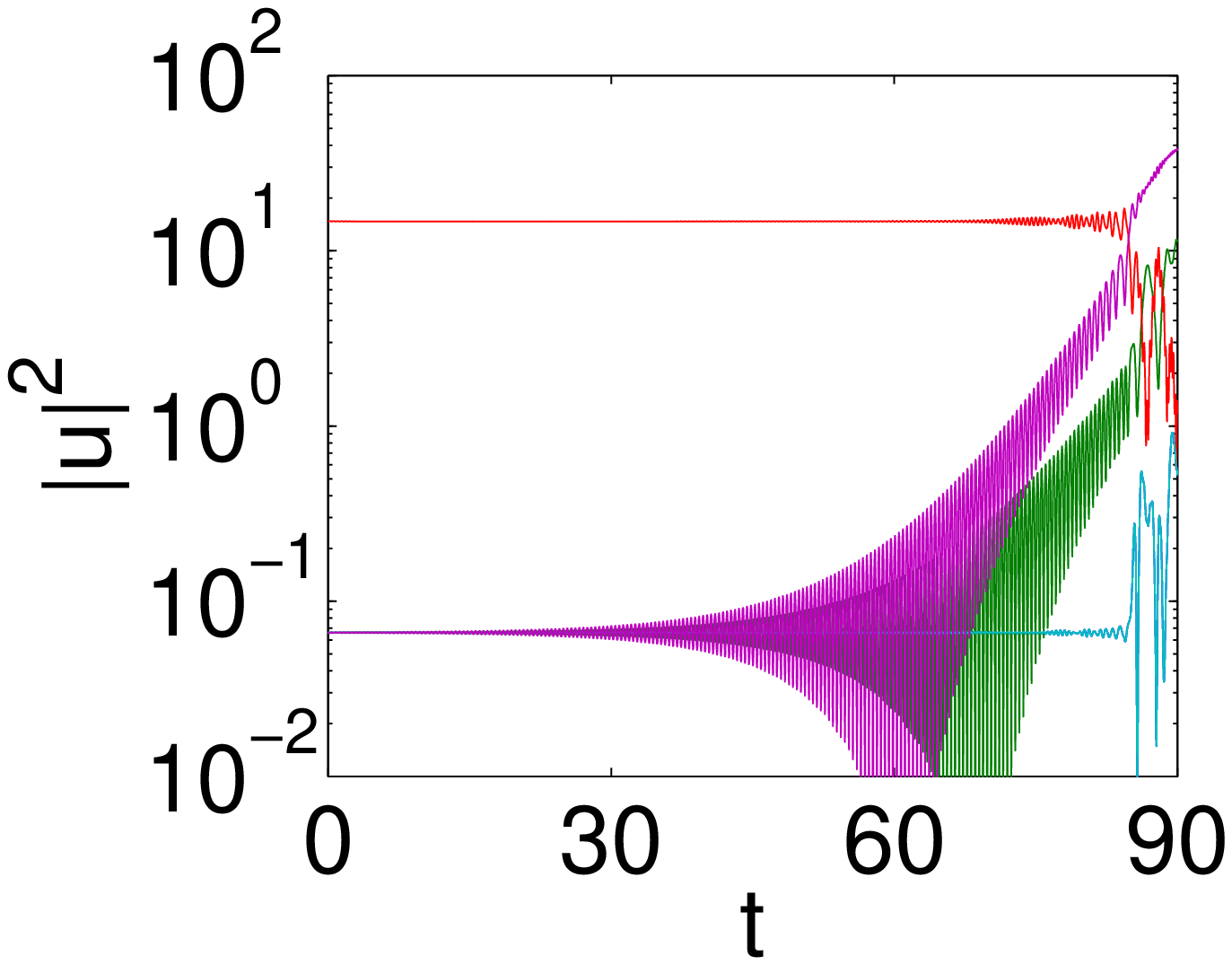}}}
\subfigure[\ red
crosses branch]{\scalebox{0.4}{\includegraphics{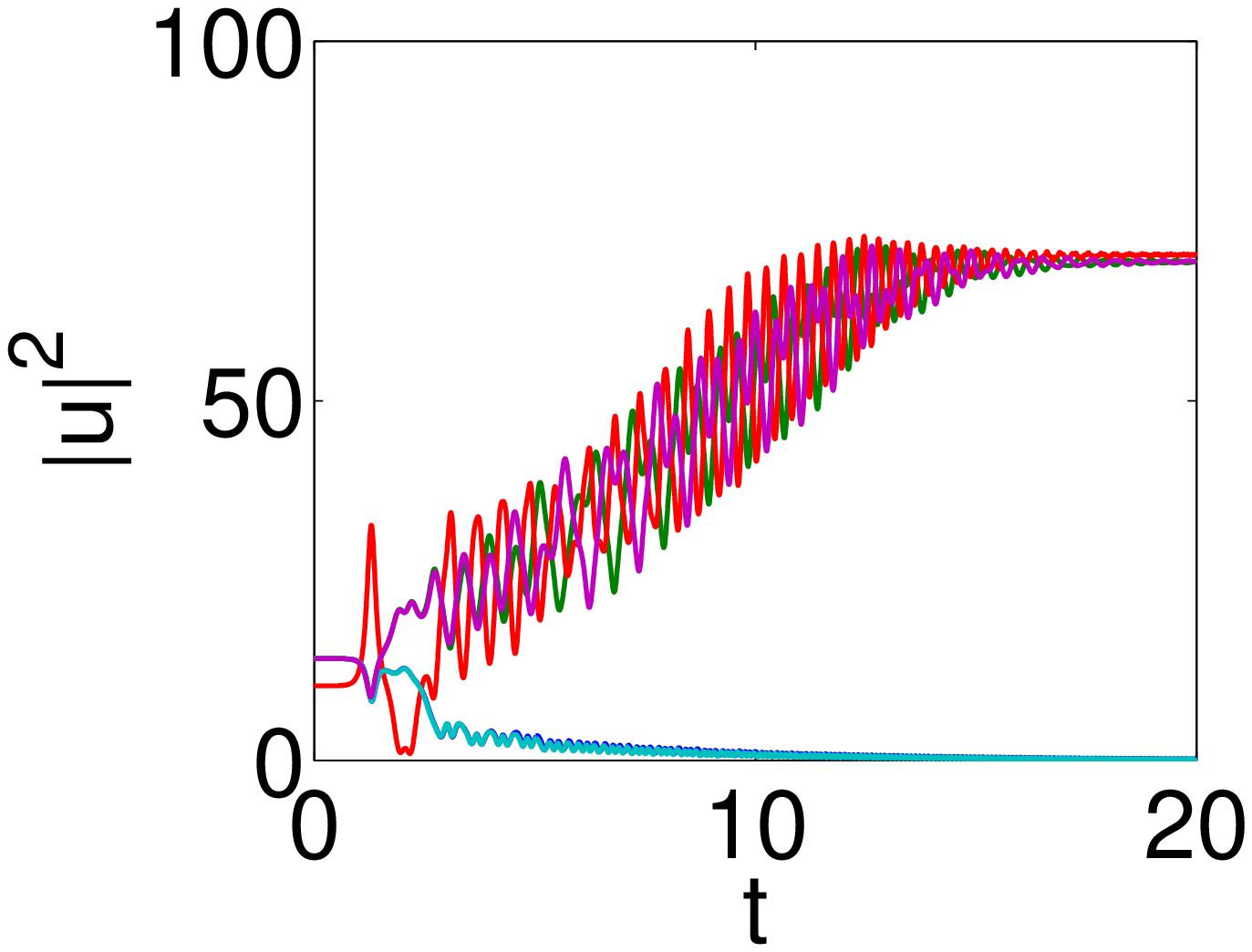}}}
\subfigure[\ black squares
branch]{\scalebox{0.4}{\includegraphics{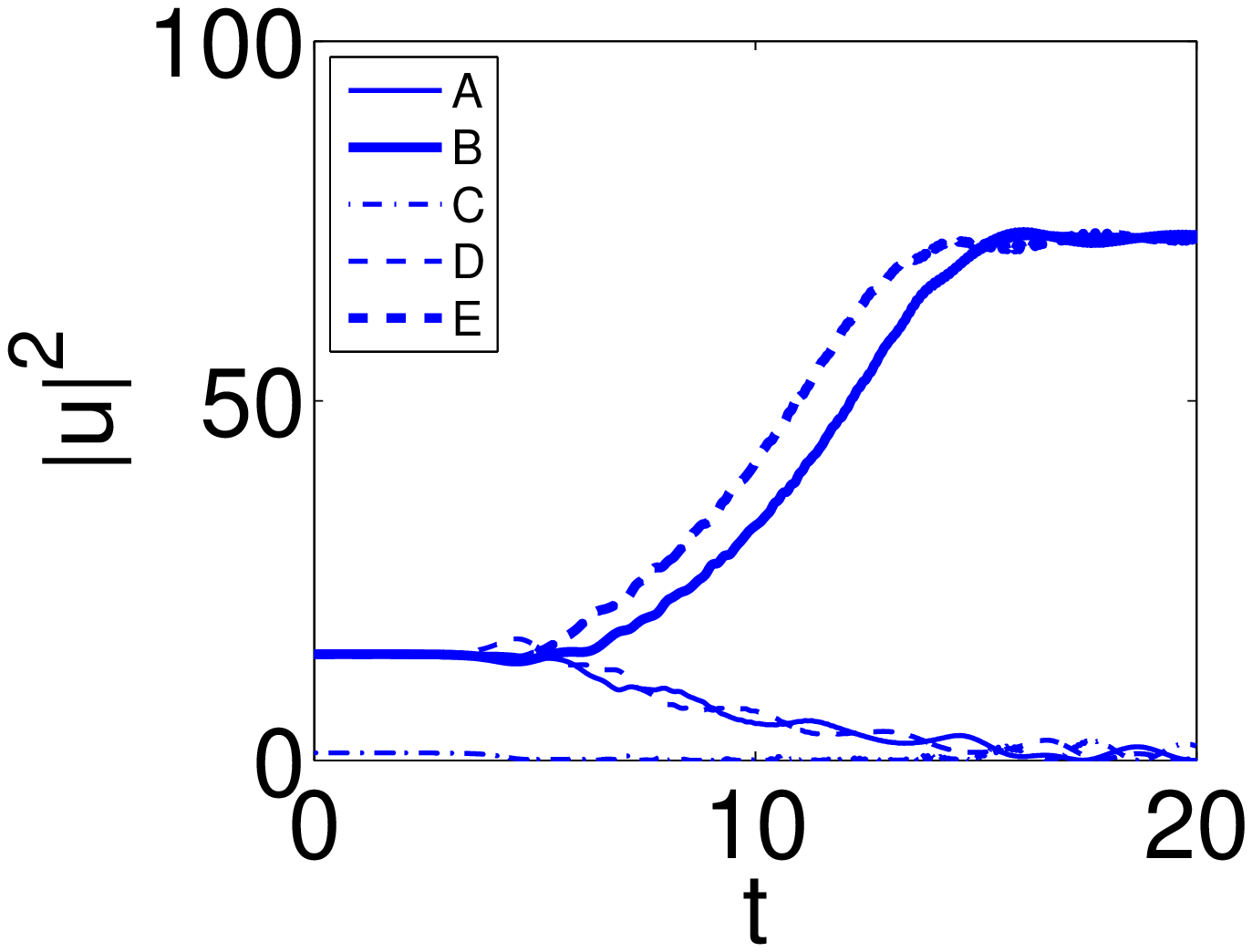}}}
\subfigure[\
green stars branch]{\scalebox{0.4}{\includegraphics{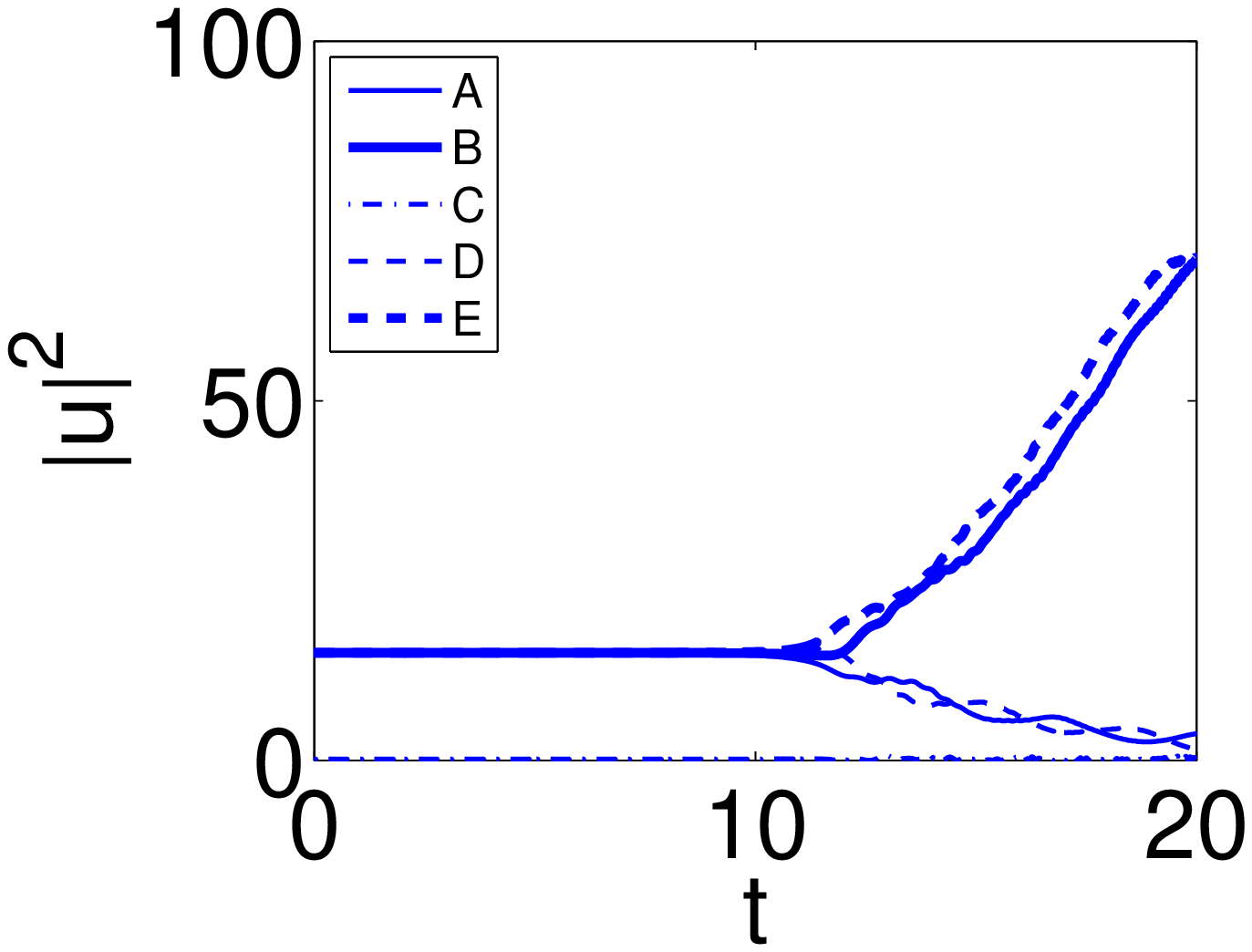}}}
\subfigure[\ magenta diamonds
branch]{\scalebox{0.4}{\includegraphics{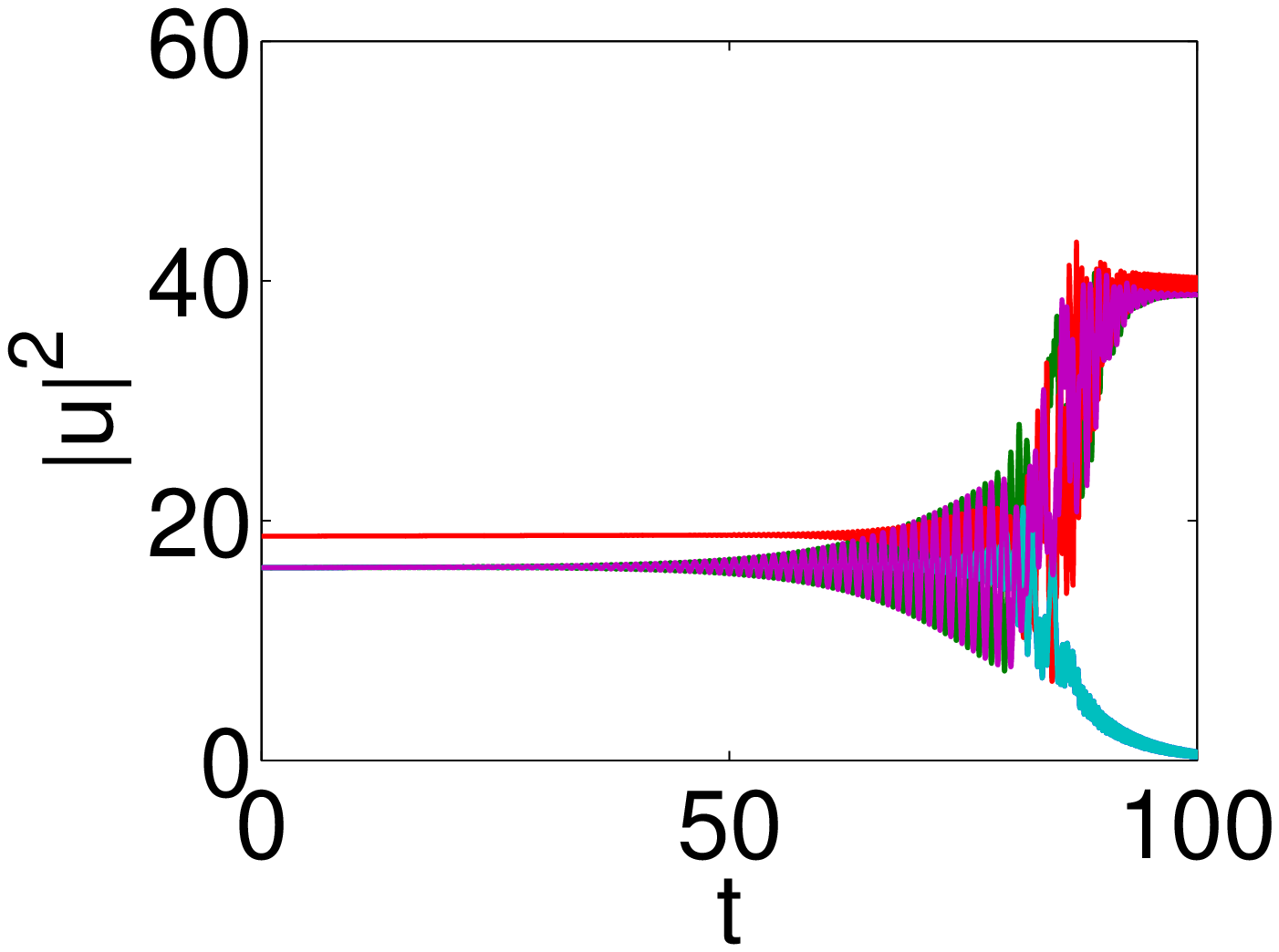}}}
\caption{(Color online) The perturbed evolution for solutions belonging to
different branches from Figs.~\protect\ref{figpmzpm1} and Fig.~\protect\ref%
{figpmzpm2}, at $\protect\gamma =0.1$. In panel (a), the amplitudes at the
different sites of plaquette (d) from Fig. \protect\ref{modes} (A,B,C,D,E)
are depicted as follows. A: the line around $10^{-1}$; B: the right one of
the two triangle-like (oscillating) curves; C: the line around $10^{1}$; D:
overlapped by A; E: the left one of the two triangle like curves. In panel
(b), the amplitudes at sites A and D overlap with each other and correspond
to the bottom curve which tends to 0, while the amplitudes at sites B, C, E
eventually grow to a large value. Panels (c) and (d) represent the dynamical
effect of the gain at sites B and E, and loss at sites A and D, while the
curve for the amplitude at site C remains very close to zero. (e) A and D
overlap with each other and correspond to the bottom curve, which tends to
0; B, C, E eventually grow to values $\simeq 40$. B and D overlap with each
other and C starts a little higher than those two.}
\label{stabpmzpm}
\end{figure}

The dynamics of the solutions belonging to these branches is shown in Fig.~%
\ref{stabpmzpm}. For the branches depicted by black squares and green stars
(recall that they disappear through the collision and the first saddle-center
bifurcation at $\gamma =0.13$), the perturbed evolution is fairly simple:
the amplitudes grow at the gain-carrying sites and decay at the lossy ones,
while the central passive site (C) stays almost at zero amplitude. For the
other branches, the amplitudes also grow at the two gain-carrying sites and
decay at the lossy elements, while the passive site may be drawn to either
the growth or decay.

\begin{figure}[htp]
\label{figpmzpm2}\scalebox{0.4}{\includegraphics{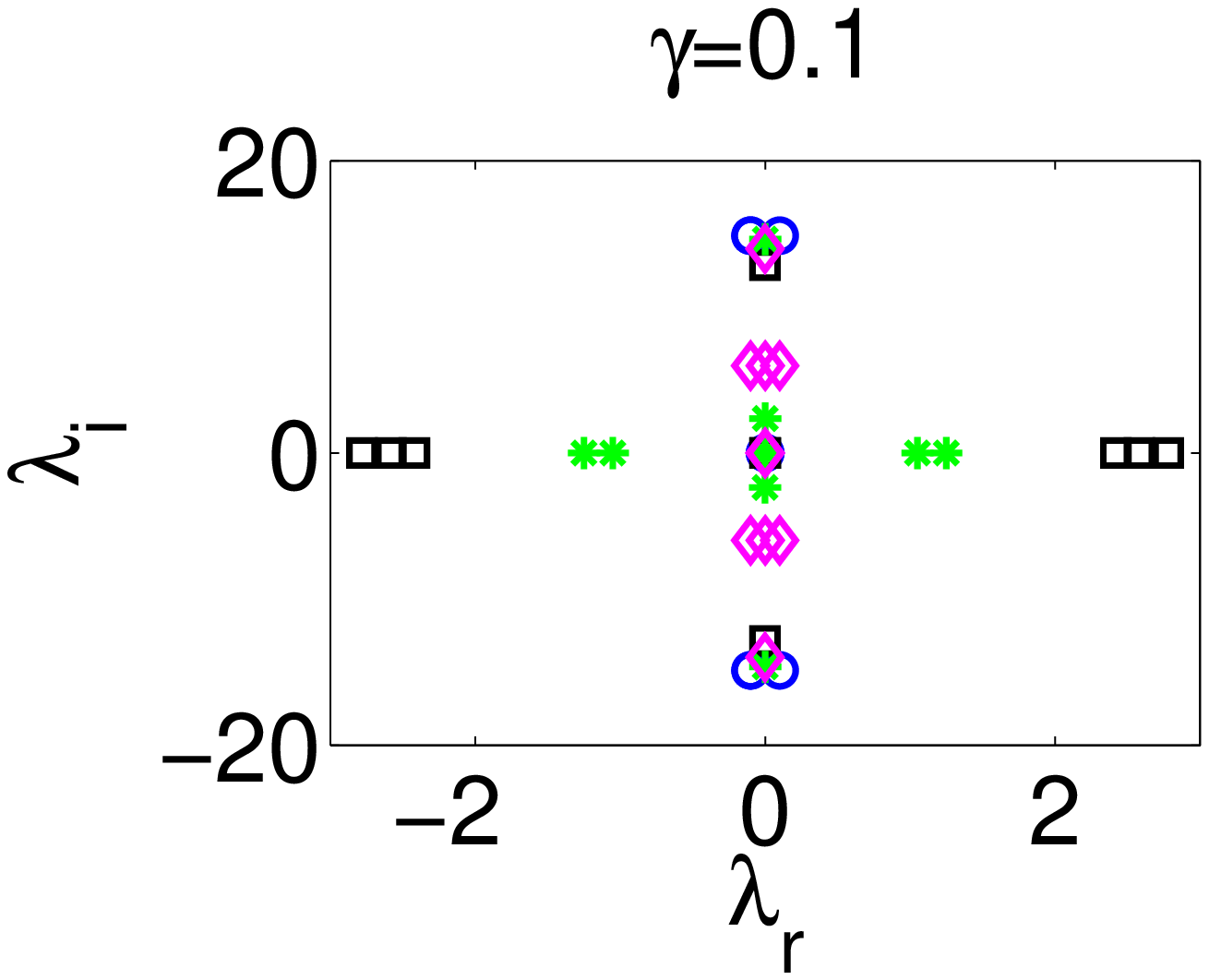}} %
\scalebox{0.4}{\includegraphics{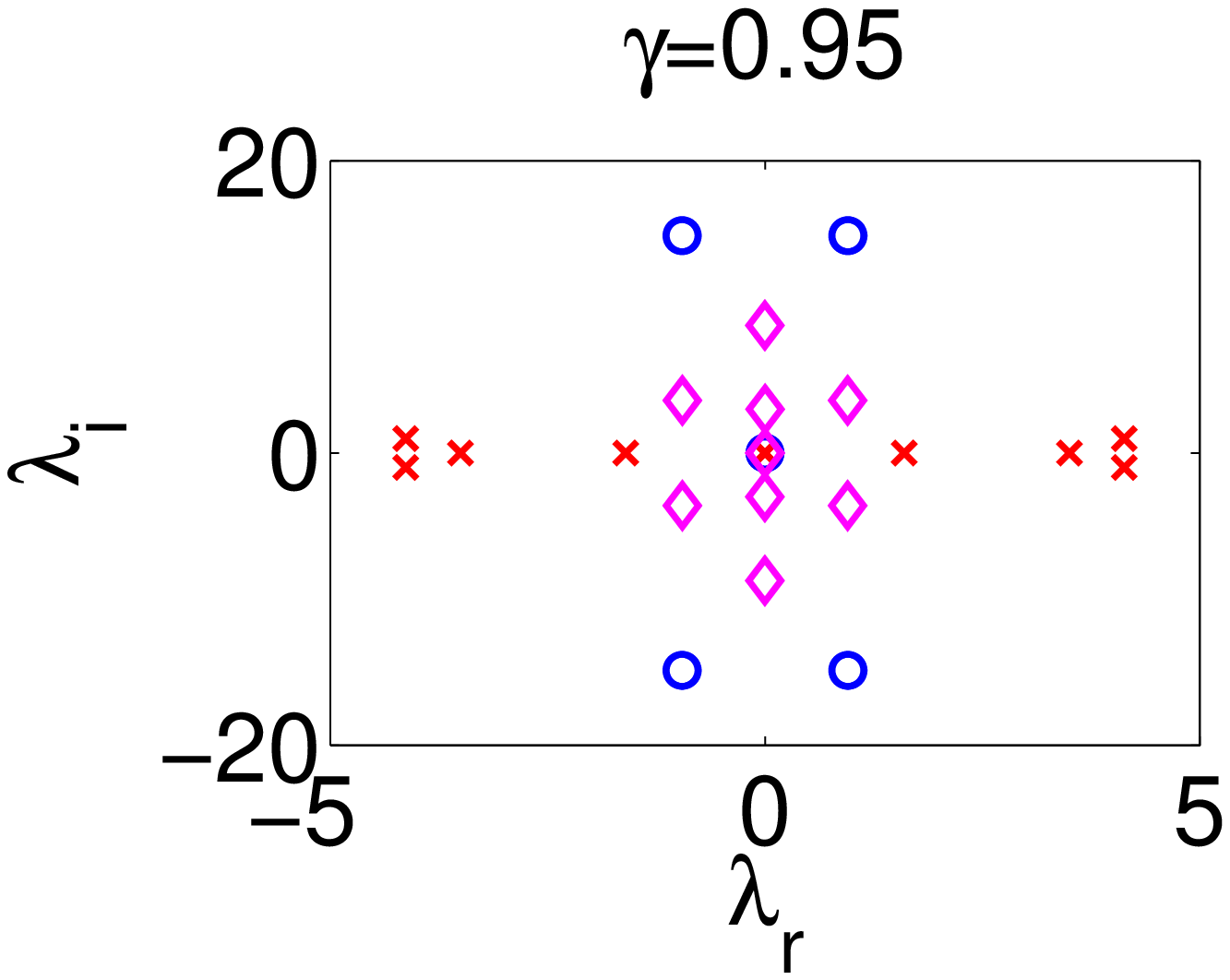}}
\caption{(Color online) Case examples of the spectral planes of the
linear-stability eigenvalues for the different solution branches shown in
the previous figure, for $G=15$, $k=1$, and $\protect\gamma =0.1$ and $0.95$
(from left to right).}
%The circled blue one always has a quartet of complex eigenvalues.
%The crossed red one (not shown in the left panel due to its large scale)
%has four pairs real eigenvalues, two of which collide into a complex
%quartet at $\gamma=0.92$.
%The diamonded magenta one always has a complex quartet and two pairs of purely
%imaginary eigenvalues.
%The squared black one always has three pairs of real and one pair of purely
%imaginary eigenvalues.
%The stared green one always has two pairs of real and two pairs of purely
%imaginary eigenvalues.
%The diamonded magenta and the crossed red terminate at the same point when $\gamma=1$;
%the squared black and stared green terminate at the same point when $\gamma=0.13$;
%the circled blue does not terminate.}
\end{figure}

\section{Conclusions and Future Challenges\label{conclu}}

In the present work, we have proposed generalizations of the one-dimensional
$\mathcal{PT}$-symmetric nonlinear oligomers into two-dimensional
plaquettes, which may be subsequently used as fundamental building blocks
for the construction of $\mathcal{PT}$-symmetric two-dimensional lattices.
In this context, we have introduced four basic types of plaquettes, three of
which in the form of four-site squares. The final one was in the form of the
five-site cross, motivated by earlier works on cross-shaped (alias rhombic
or site-centered) vortex solitons in the discrete nonlinear Schr{\"{o}}%
dinger equation. Our analysis was restricted to modes which could be found
in the analytical form, while their stability against small perturbations
was analyzed by means of numerical methods. Even within the framework of
this restriction, many effects have been found, starting from the existence
of solution branches that terminate at the critical points of the respective
linear $\mathcal{PT}$-symmetric systems --- e.g., in the settings
corresponding to plaquettes (a) and (c) in Fig. \ref{modes}. The bifurcation
responsible for the termination of the pair of branches may take a complex
degenerate form [such as the one in the case of setting (a)]. Other branches
were found too, that continue to exist, due to the nonlinearity, past the
critical points of the underlying linear systems. In addition, we have
identified cases [like the gain-loss alternating pattern (b) or the cross
plaquette of type (d)] when the $\mathcal{PT}$ symmetry is broken
immediately after the introduction of the gain-loss pattern. The spectral
stability of the different configurations was examined. Most frequently, the
stationary modes are unstable, although stable branches were found too
[e.g., in settings (a) and (c)]. We have also studied the perturbed dynamics
of the modes. The evolution of unstable ones typically leads to the growth
of the amplitudes at the gain-carrying sites and decay at the lossy ones. It
was interesting to observe that the passive sites, without gain or loss,
might be tipped towards growth or decay, depending on the particular
solution (and possibly on specific initial conditions).

The next relevant step of the analysis may be to search for more
sophisticated stationary modes (that plausibly cannot be found in an
analytical form), produced by the \textit{symmetry breaking} of
the simplest modes considered in this work, cf. Ref. \cite{miron}. The
difference of such modes from the $\mathcal{PT}$-symmetric ones considered
in the present work is the fact that modes with the unbroken symmetry form a
continuous family of solutions, with energy $E$ depending on the solution's
amplitude, see Eq. (\ref{E}). This feature, which is generic to conservative
nonlinear systems, is shared by $\mathcal{PT}$-symmetric ones, due to the
\textquotedblleft automatic" balance between the separated gain and loss. On
the other hand, the breaking of the symmetry gives rise to the
typical behavior of systems with competing, but not explicitly balanced,
gain and loss, which generate a single or several \textit{attractors}, i.e.,
\emph{isolated} solutions with a single or several values of the energy,
rather than a continuous family. A paradigmatic example of the difference
between continuous families of solutions in conservative models and isolated
attractors in their (weakly) dissipative counterparts is the transition from
the continuous family of solitons in the usual NLSE to a pair of isolated
soliton solutions, one of which is an attractor (and the other is an
unstable solution playing the role of the separatrix between attraction
basins,  the stable soliton and the stable zero solution) in the complex
Ginzburg-Landau equation, produced by the addition of the cubic-quintic
combination of small dissipation and gain terms to the NLSE \cite{PhysicaD}.
As concerns the systems considered in the present work, in the context of
the breaking of the $\mathcal{PT}$ symmetry it may also be
relevant to introduce a more general nonlinearity, which includes $\mathcal{%
PT}$-balanced cubic gain and loss terms, in addition to their linear
counterparts (cf. Refs. \cite{konorecent} and \cite{miron}).
Nevertheless, it should also be noted that the issue of potential
existence of isolated solutions versus branches of solutions in
 $\mathcal{PT}$-symmetric systems is already starting to be addressed
in the relevant literature (including in plaquette-type configurations),
as in the very recent work of~\cite{konorecent4}.

Moreover, the present work may pave the way to further considerations of
two-dimensional $\mathcal{PT}$-symmetric lattice systems, and even
three-dimensional ones. In this context, the natural generalization is to
construct periodic two-dimensional lattices of the building blocks presented
here, and to identify counterparts of the modes reported here in the
infinite lattices, along with new modes which may exist in that case. On the
other hand, in the three-dimensional realm, the first step that needs to be
completed would consist of the examination of a $\mathcal{PT}$-symmetric
cube composed of eight sites, and the nonlinear modes that it can support.
This, in turn, may be a preamble towards constructing full three-dimensional
$\mathcal{PT}$-symmetric lattices. These topics are under present
consideration and will be reported elsewhere.

\section*{Acknowledgments}

UG thanks Holger Cartarius and Eva-Maria Graefe for useful discussions.
PGK gratefully acknowledges support from the National Science Foundation
under grant DMS-0806762 and CMMI-1000337, as well as from the Alexander von
Humboldt Foundation and the Alexander S. Onassis Public Benefit Foundation.
PGK and BAM also acknowledge support from the Binational Science Foundation
under grant 2010239.


\section*{References}

\begin{thebibliography}{99}
\bibitem{bend} C. M. Bender and S. Boettcher, Phys. Rev. Lett. \textbf{80},
5243 (1998); C. M. Bender, S. Boettcher and P. N. Meisinger, J. Math. Phys.
\textbf{40}, 2201 (1999). %;
%C.M. Bender, Rep. Prog. Phys. {\bf 70}, 947 (2007).

\bibitem{christo1} Z. H. Musslimani, K. G. Makris, R. El-Ganainy and D. N.
Christodoulides, Phys. Rev. Lett. \textbf{100}, 030402 (2008); K. G. Makris,
R. El-Ganainy, D. N. Christodoulides and Z. H. Musslimani, Phys. Rev. A
\textbf{81}, 063807 (2010).

\bibitem{haifa-prl-2008} S. Klaiman, U. G\"unther, N. Moiseyev, Phys. Rev.
Lett. \textbf{101}, 080402 (2008).

\bibitem{salamo} A. Guo, G. J. Salamo, D. Duchesne, R. Morandotti, M.
Volatier-Ravat, V. Aimez, G. A. Siviloglou and D. N. Christodoulides, Phys.
Rev. Lett. \textbf{103}, 093902 (2009).

\bibitem{kip} C. E. R{\"{u}}ter, K. G. Makris, R. El-Ganainy, D. N.
Christodoulides, M. Segev, D. Kip, Nature Phys. \textbf{6}, 192 (2010).

\bibitem{Winful} B. A. Malomed and H. G. Winful, Phys. Rev. E \textbf{53},
5365 (1996); H. Sakaguchi and B. A. Malomed, Physica D \textbf{147}, 273
(2000); W. J. Firth and P. V. Paulau, Eur. Phys. J. D \textbf{59}, 13
(2010); P. V. Paulau, D. Gomila, P. Colet, N. A. Loiko, N. N. Rosanov, T.
Ackemann, and W. J. Firth, Opt. Express \textbf{18}, 8859 (2010); A. Marini,
D. V. Skryabin, and B. A. Malomed, \textit{ibid}. \textbf{19}, 6616 (2011);
P. V. Paulau, D. Gomila, P. Colet, B. A. Malomed, and W. J. Firth, Phys.
Rev. E \textbf{84}, 036213 (2011).

\bibitem{Atai} J. Atai and B. A. Malomed, Phys. Rev. E \textbf{54}, 4371
(1996).

\bibitem{Chaos} B. A. Malomed, Chaos \textbf{17}, 037117 (2007).

\bibitem{tsampikos_recent} J. Schindler, A. Li, M. C. Zheng, F. M. Ellis and
T. Kottos, Phys. Rev. A \textbf{84}, 040101 (2011).

\bibitem{tsampikos_recent2} H. Ramezani, J. Schindler, F. M. Ellis, U.
G\"unther, and T. Kottos, arXiv:1205.1847.

\bibitem{darmstadt-pt} S. Bittner, B. Dietz, U. G\"unther, H. L. Harney, M.
Miski-Oglu, A. Richter, and F. Sch\"afer, Phys. Rev. Lett. \textbf{108}, 024101 (2012).

\bibitem{cart-wun2012} H. Cartarius and G. Wunner, arXiv:1203.1885 (to be published in the present special issue).

\bibitem{pgk} K. Li and P. G. Kevrekidis Phys. Rev. E \textbf{83}, 066608
(2011).

\bibitem{kot1} H. Ramezani, T. Kottos, R. El-Ganainy and D. N.
Christodoulides, Phys. Rev. A \textbf{82}, 043803 (2010).

\bibitem{sukh1} A.A. Sukhorukov, Z. Xu and Yu. S. Kivshar, Phys. Rev. A
\textbf{82}, 043818 (2010).

\bibitem{kot2} M. C. Zheng, D. N. Christodoulides, R. Fleischmann and T.
Kottos, Phys. Rev. A \textbf{82}, 010103(R) (2010).

\bibitem{grae1} E. M. Graefe, H. J. Korsch and A. E. Niederle, Phys. Rev.
Lett. \textbf{101}, 150408 (2008).

\bibitem{grae2} E. M. Graefe, H. J. Korsch and A. E. Niederle, Phys. Rev. A
\textbf{82}, 013629 (2010).

\bibitem{kot3} Z. Lin, H. Ramezani, T. Eichelkraut, T. Kottos, H. Cao and D.
N. Christodoulides, Phys. Rev. Lett. \textbf{106}, 213901 (2011).

%\bibitem{pgk} K. Li and P. G. Kevrekidis
%Phys. Rev. E {\bf 83}, 066608 (2011)

\bibitem{dmitriev1} S. V. Dmitriev, S. V. Suchkov, A.A. Sukhorukov, and Yu.
S. Kivshar, Phys. Rev. A \textbf{84}, 013833 (2011)

\bibitem{dmitriev2} S. V. Suchkov, B.A. Malomed, S. V. Dmitriev and Yu. S.
Kivshar, Phys. Rev. E 84, 046609 (2011); S. V. Suchkov, S. V. Dmitriev, B.
A. Malomed, and Y. S. Kivshar, Phys. Rev. A \textbf{85}, 033835 (2012).

\bibitem{miron} A. E. Miroshnichenko, B.A. Malomed, and Yu. S. Kivshar Phys.
Rev. A \textbf{84}, 012123 (2011).

\bibitem{konorecent} F. Kh. Abdullaev, Y. V. Kartashov, V. V. Konotop and D.
A. Zezyulin, Phys. Rev. A \textbf{83}, 041805 (2011)

\bibitem{konorecent2} D. A. Zezyulin, Y. V. Kartashov, and V. V. Konotop,
Europhys. Lett. {\bf 96}, 64003 (2011).
%arXiv:1111.0898.

\bibitem{YJH} Y. He, X. Zhu, D. Mihalache, J. Liu, and Z. Chen, Phys. Rev. A
\textbf{85}, 013831 (2012).

\bibitem{Yang} S. Nixon, L. Ge, and J. Yang, Phys. Rev. A \textbf{85},
023822 (2012).

\bibitem{Review} F. Lederer, G. I. Stegeman, D. N. Christodoulides, G.
Assanto, M. Segev, and Y. Silberberg, Phys. Rep. \textbf{463}, 1 (2008).

\bibitem{pgk_book} P. G. Kevrekidis \textit{The discrete nonlinear Schr{\"{o}%
}dinger equation: Mathematical Analysis, Numerical Computations and Physical
Perspectives}, Springer-Verlag (Heidelberg, 2009).

\bibitem{malomed} B. A. Malomed and P. G. Kevrekidis Phys. Rev. E \textbf{64}%
, 026601 (2001).

\bibitem{pelin} D. E. Pelinovsky, P. G. Kevrekidis, D. J. Frantzeskakis,
Phys. D \textbf{212}, 1 (2005).

\bibitem{moti} J. W. Fleischer, G. Bartal, O. Cohen, O. Manela, M. Segev, J.
Hudock, and D. N. Christodoulides Phys. Rev. Lett. \textbf{92}, 123904
(2004).

\bibitem{yuri} D. N. Neshev, T. J. Alexander, E. A. Ostrovskaya, Yu. S.
Kivshar, H. Martin, I. Makasyuk, and Z. Chen, Phys. Rev. Lett. \textbf{92},
123903 (2004).

\bibitem{konorecent4} D. A. Zezyulin and V. V. Konotop,
Phys. Rev. Lett. {\bf 108}, 213906 (2012).
%arXiv:1202.3652.

\bibitem{grae3} E. M. Graefe, J. Phys. A (contribution to the present special issue).

\bibitem{nonlin-1} V. I. Arnold, Comm. Pure Appl. Math. \textbf{29}, 557
(1976).

\bibitem{nonlin-2} J. Moser, Comm. Pure Appl. Math. \textbf{29} 727 (1976).

\bibitem{nonlin-3} M. Golubitsky and I. Stewart, Arch. Rat. Mech. Anal.
\textbf{87}, 107 (1985).

\bibitem{nonlin-4} C. Elphick, E. Tirapegui, M. E. Brachet, P. Coullet, and
G. Iooss, Physica D \textbf{29}, 95 (1987).

\bibitem{nonlin-5} J. Montaldi, M. Roberts and I. Stewart, Nonlinearity
\textbf{3}, 695 (1990).

\bibitem{nonlin-6} G. M. Chechin and V. P. Sakhnenko, Physica D \textbf{117}%
, 43 (1998).

\bibitem{nonlin-7} A. Ferrando, M. Zacar\'{e}s, P. Andrees, P. Fernandez de
Cordoba and J. A. Monsoriu, Optics Express \textbf{13}, 1073 (2005).

\bibitem{nonlin-8} M. Zacar\'{e}s, M. Arevalillo-Herraez and S. Abraham,
Comp. Phys. Comm. \textbf{181}, 35 (2010).

\bibitem{wigner-book} E. Wigner, Group Theory and Its Application to Quantum
Mechanics of Atomic Spectra, (Academic Press, 1959).

\bibitem{Arnold-degen} V. I. Arnold, Russ. Math. Surv. \textbf{26}, \# 2, 29
(1972).

\bibitem{GS2005}U. G\"unther and F. Stefani, Czech. J. Phys. {\bf 55},  1099-1106 (2005); math-ph/0506021.

\bibitem{heissEP3} W. D. Heiss, J. Phys. A: Math. Theor. {\bf 41}, 244010 (2008).

\bibitem{GGKN2008} E. M. Graefe, U. G\"unther, H. J. Korsch, and A. E. Niederle, J. Phys. A: Math. Theor. {\bf 41}, 255206 (2008).

\bibitem{grae4} G. Demange and E. M. Graefe, J. Phys. A: Math. Theor. {\bf 45}, 025303 (2012).

\bibitem{znojil-2001} M. Znojil, Rendic. Circ. Mat. Palermo, Ser. II, Suppl. {\bf 72} (2004), 211 - 218, math-ph/0104012.
\bibitem{japa-2001} G. S. Japaridze, 	J. Phys. A: Math. Theor. {\bf 35}, 1709-1718 (2002), quant-ph/0104077.
\bibitem{ali-2001} A. Mostafazadeh, J. Math. Phys. {\bf 43}, 205-214 (2002), math-ph/0107001.
\bibitem{cmb-2002} C. M. Bender, D. C. Brody,  and H. F. Jones, Phys. Rev. Lett. {\bf 89}, 270401 (2002), quant-ph/0208076.


%\bibitem{complex-linearization} P.G. Kevrekidis,
%The discrete nonlinear Schr{\"o}dinger equation: mathematical
%analysis, numerical computation and physical perspectives, (Springer-Verlag,
%Heidelberg 2009).

%\bibitem{konorecent4} D. A. Zezyulin and V. V. Konotop, arXiv:1202.3652.

\bibitem{PhysicaD} B. A. Malomed, Physica D \textbf{29}, 155 (1987).
\end{thebibliography}
\end{document}